\documentclass[11pt]{article}

\pdfoutput=1
\usepackage{color}
\usepackage{amsmath,amssymb,hyperref,bm,url}
\usepackage{slashed,relsize}
\usepackage{graphicx}
\usepackage[T1]{fontenc}
\usepackage{cite}
\usepackage{hyperref}
\def\beq{\begin{equation}\displaystyle}
\def\eeq{\end{equation}}
\def\bea{\begin{eqnarray}\displaystyle}
\def\eea{\end{eqnarray}}

\title{\bf{Learning New Physics from a Machine}}

\date{\today}

\author{Raffaele Tito D'Agnolo$^{1}$,\, Andrea Wulzer$^{2,3,4}$\\[2pt]
{\small\emph{$^{1}$ {SLAC National Accelerator Laboratory, 2575 Sand Hill Road, Menlo Park, CA, 94025, USA.}}}\\
{\small\emph{$^{2}$CERN, Theoretical Physics Department, Geneva, Switzerland}}\\
{\small\emph{$^{3}$ Institut de Th\'eorie des Ph\'enomenes Physiques, EPFL, Lausanne, Switzerland}}\\
{\small\emph{$^{4}$Dipartimento di Fisica e Astronomia, Universit\`a di Padova and}}\\
{\small\emph{INFN, Sezione di Padova, via Marzolo 8, I-35131 Padova, Italy}}
}
\begin{document}
\baselineskip=13pt
\maketitle

\begin{abstract}
We propose using neural networks to detect data departures from a given reference model, with no prior bias on the nature of the new physics responsible for the discrepancy. The virtues of neural networks as unbiased function approximants make them particularly suited for this task. An algorithm that implements this idea is constructed, as a straightforward application of the likelihood-ratio hypothesis test. The algorithm compares observations with an auxiliary set of reference-distributed events, possibly obtained with a Monte Carlo event generator. It returns a $p$-value, which measures the compatibility of the reference model with the data. It also identifies the most discrepant phase-space region of the data set, to be selected for further investigation. The most interesting potential applications are model-independent new physics searches, although our approach could also be used to compare the theoretical predictions of different Monte Carlo event generators, or for data validation algorithms.

In this work we study the performance of our algorithm on a few simple examples. The results confirm the model-independence of the approach, namely that it displays good sensitivity to a variety of putative signals. Furthermore, we show that the reach does not depend much on whether a favorable signal region is selected based on prior expectations. We identify directions for improvement towards applications to real experimental data sets.
\end{abstract}


\newpage
\section{Introduction}\label{sec:intro}

Today in fundamental physics we have at our disposal powerful theoretical models. They are in principle able to describe the outcome of all present and near-future experiments. In high-energy physics and cosmology these model are the Standard Model (SM) and $\Lambda$CDM, respectively. In the following we call them reference models. It is technically possible for the reference models to describe all present and future data, but that does not mean that they will. Future experiments will be able to explore phenomena that we have never observed before, or to measure known phenomena with unprecedented accuracy. Furthermore, we are convinced that new physics (i.e., physical laws that are not yet established) exists, because of the open problems of the reference models. Searching for new physics, which concretely means searching for discrepancies between the data and the reference model, is the absolute priority of our field.

In general the problem can be phrased in terms of many repeated measurements ${\mathcal{D}}=\{x_i\}$ (called events in high-energy physics) of a multi-dimensional random variable $x$. The statistical distribution for $x$ can be predicted on the basis of the physical laws that constitute the reference model. The goal is to test the reference model distribution against the actual data. Several strategies exist to carry out this test. However the vast majority of them are not suited to discover discrepancies because of the nature of the problem at hand. The main challenge stems from the fact that the true underlying data distribution, possibly including new physics effects, will be ``similar'' to the reference one. We expect this because of existing constraints on new physics. Notice that ``similar'' does not mean that the effect of new physics cannot be large. However if it is large it will be localized in a low-probability region of the space of observations where only a a small fraction of the events is present. Alternatively the effect can be spread in a large region of the $x$ space, but in this case it will be a small modification of the reference distribution. Essentially the problem is that our prior knowledge suggests that the vast majority of the collected events will agree with the reference model. At the same time this prior knowledge is insufficient to know where to look for discrepancies. 

The most widely employed approach to the problem is to search for specific new physics models. In any such model, one can identify a priori the subset of data where large departures from the reference model should be concentrated, or know how to exploit small, correlated deviations across the data set. Once a specific new physics model or a set of models are specified, one constructs hypothesis tests using standard techniques (see \cite{Patrignani:2016xqp} for a concise review). The clear advantage of this approach is that it is physically informative even if the compatibility of the data with the reference model is confirmed. The disadvantage is that a statistical test which is designed to be sensitive to one specific hypothesis is typically insensitive to data departures of different nature. This substantiates the widespread concern that we might not be able to discover new physics, even if present in the data, because it does not belong to the class of hypothetical models that we are searching for. 
 
Motivated by the above observation, a number of attempts have been made \cite{Choudalakis:2011qn, Abbott:2000fb, Abbott:2000gx, Aktas:2004pz, Aaron:2008aa, Asadi:2017qon, Aaltonen:2007dg, Aaltonen:2008vt, CMS:2008gya, CMS:2011fra, ATLAS:2017irs, ATLAS:2012qna, ATLAS:2014sxa} to construct model-independent new physics search strategies. However it is important to remark that a model-independent hypothesis test is an ill-defined concept in statistics. Testing one hypothesis unavoidably requires an alternative (in general composite) hypothesis to compare with. Technically what is needed is a set of alternative hypothetical distributions, depending on free parameters, which are also called an alternative probability ``model'' in statistics. In physics instead a model is a set of physical laws that allows to predict these distributions. Therefore we call a search strategy model-independent (in the physics sense) when the alternative distributions do not follow from a physical model, but are selected with other criteria. The most important criterion is flexibility, namely the ability of the distributions to adapt themselves, for an appropriate choice of the free parameters, to the true underlying data distribution. This will ensure sensitivity to a large variety of new physics scenarios, including those that are not predicted by any of the models that have been constructed until now. The idea behind the present paper is to use artificial neural networks to parameterize the alternative distributions.

Neural networks are increasingly important tools in high energy physics. Applications include jet physics \cite{deOliveira:2015xxd, Schwartzman:2016jqu, Kagan:2016wnu,	Larkoski:2017jix, Louppe:2017ipp, Shimmin:2017mfk, Baldi:2016fql, Guest:2016iqz, Almeida:2015jua, Barnard:2016qma, Kasieczka:2017nvn, Butter:2017cot, Datta:2017rhs, Datta:2017lxt, Fraser:2018ieu, Andreassen:2018apy, Macaluso:2018tck, ATLAS:2017jiz, CMS-DP-2017-013, ATL-PHYS-PUB-2017-013, ATL-PHYS-PUB-2017-004, CMS-DP-2017-005, ATL-PHYS-PUB-2017-003, ATL-PHYS-PUB-2017-017, CMS-DP-2017-027}, new physics searches~\cite{Baldi:2016fzo, Chang:2017kvc, Cohen:2017exh, Brehmer:2018eca, Brehmer:2018kdj, Brehmer:2018hga, Roxlo:2018adx, Collins:2018epr}, detector simulation~\cite{Paganini:2017dwg,deOliveira:2017rwa,Paganini:2017hrr} and the NNPDF fit to parton distribution functions \cite{Ball:2014uwa}, where they have been successfully applied for a long time \cite{Forte:2002fg}. The main reason for their success is precisely their virtues as efficient and unbiased approximants \cite{Cybenko1989,KREINOVICH1991381,Hecht-Nielsen:1992:TBN:140639.140643,HORNIK1989359,2016arXiv161004161L,DBLP:journals/corr/PoggioMRML16,DBLP:journals/corr/Bach14,2017arXiv171208688M}. They are often introduced as a convenient alternative to piecewise constant functions (histograms) for the fit to distributions~\cite{bishop:2006:PRML,Goodfellow-et-al-2016,Haykin1998}. Employing them to parametrize alternative distributions for model-independent new physics searches is thus a highly motivated attempt. To the best of our knowledge this possibility has not been previously discussed. Most applications of neural networks to new physics searches aim at enhancing the sensitivity to pre-specified models of the resonant or non-resonant type. Using machine learning techniques for model-independent new physics searches has been proposed in \cite{Kuusela:2011aa}, however Gaussian mixture models are employed rather than neural networks and the overall strategy is quite different from ours. Ref.~\cite{Collins:2018epr} uses neural networks, but with the purpose of enhancing the sensitivity to resonant bumps that emerge in a pre-specified kinematical variable. What we do is conceptually very similar to anomaly detection, where neural networks are already employed extensively. However the purpose of anomaly detection is to identify rare events in the data sample. Our purpose is instead to find an anomalous behavior, relative to the reference model, of the entire data sample.

The paper is organized as follows. In section~\ref{sec:bi} we introduce the conceptual foundations of our approach, explaining in detail the advantages of using neural networks for model-independent new physics searches. We will see that our strategy is a straightforward application of maximum likelihood estimation and likelihood-ratio hypothesis testing, which are easily turned into a neural network training problem as shown in section~\ref{sec:alg}. In section~\ref{numexp} we perform several numerical experiments to illustrate the virtues of our algorithm and its limitations. A slightly different perspective on the foundations of our method, which offers more flexibility in the implementation, is discussed in section~\ref{sec:altloss}. Our conclusions are reported in section~\ref{conc}, together with a discussion of other possible applications. These are comparisons of different Monte Carlo generators and data validation algorithms.

\section{Conceptual Foundations}\label{sec:bi}

Consider repeated measurements ${\mathcal{D}}=\{x_i\}$, $i=1,\ldots,{\mathcal{N}}_{{\mathcal{D}}}$ of a $d$-dimensional random variable $x$, and let $n(x|{\rm{R}})$ be its differential distribution as predicted by the reference model ``R''. Here and in what follows we denote as differential distribution the probability density function (p.d.f.) of $x$ normalized to the total number of expected events in the experiment, namely
\beq
n(x)=N\,P(x)\,,\;\;\;\;\;N=\int\hspace{-2pt}dx\,n(x)\,.
\eeq
Testing the reference model for compatibility with the observed data set ${\mathcal{D}}$ unavoidably requires comparison with an alternative hypothesis $n(x|{{\mathbf{w}}})$. In general the alternative hypothesis is composite, labeled by a number of free parameters ${{\mathbf{w}}}$. We are interested in problems where the distribution according to which the data are truly distributed is ``similar'' (in the sense specified in the Introduction) to the reference one, hence it is convenient to parametrize $n(x|{{\mathbf{w}}})$ in terms of $n(x|{\rm{R}})$. Taking also into account that $n(x|{{\mathbf{w}}})$ is necessarily positive and that we will use log-likelihood ratios for hypothesis testing, we best express it as
\beq\label{alth}
\displaystyle
n(x|{{\mathbf{w}}})=n(x|{\rm{R}})\,e^{f(x;{{{\mathbf{w}}}})}\,,
\eeq
in terms of a set of real functions ${\mathcal{F}}=\{f(x;{{{\mathbf{w}}}}), \forall\, {{{\mathbf{w}}}}\}$.

Once the set of alternative hypotheses is specified in this parametrized form, the optimal statistical test for the reference model is defined by the Neyman--Pearson construction \cite{10.2307/91247}, based on the maximum likelihood principle. The idea is to compare the reference with the best-fit distribution $n(x|{{\mathbf{\widehat{w}}}})$, obtained at the point ${{\mathbf{{w}}}}={{\mathbf{\widehat{w}}}}$ that maximizes the likelihood. This leads to the test statistic 
\beq\label{tNP}
\displaystyle
t({\mathcal{D}})=2\,\log\left[\frac{e^{-N({{\mathbf{\widehat{w}}}})}}{e^{-N({\rm{R}})}}\prod\limits_{x\in {\mathcal{D}}}\frac{n(x|{{\mathbf{\widehat{w}}}})}{n(x|{\rm{R}})}\right]=
-2\,\underset{\{{\mathbf{w}}\}}{\rm{Min}}\left[
N({{\mathbf{{w}}}})-N({\rm{R}})-\sum\limits_{x\in {\mathcal{D}}}f(x;{{{\mathbf{w}}}})
\right]
\,,
\eeq
where $N({\rm{R}})$ is the expected number of events in the reference model and $N({{\mathbf{{w}}}})$ is the expected in the alternative hypothesis, namely
\beq\label{totexp}
\displaystyle
N({{\mathbf{{w}}}})=\int\hspace{-2pt}dx\,n(x|{{\mathbf{{w}}}})=\int\hspace{-2pt}dx\,n(x|{\rm{R}})\,e^{f(x;{{{\mathbf{w}}}})}\,.
\eeq
In order to associate a probability to the value of $t$ (${t_{\rm{obs}}}$) obtained with the observed data set, the p.d.f. of $t$ in the reference hypothesis needs to be computed by repeatedly evaluating $t$ on a large sample of toy datasets. From this distribution we obtain the observed $p$-value
\beq\label{pv}
p_{\rm{obs}}=\int_{t_{\rm{obs}}}^\infty\hspace{-6pt} dt \,P(t|{\rm{R}})\,,
\eeq
defined as usual as the probability that the reference model produces a dataset that is more in tension with itself (has larger $t$) than the observed data.

The basic idea of the present paper is to parametrize the alternative hypothesis with neural networks. We take $f(x;{{{\mathbf{w}}}})$ to be fully connected neural networks, with free parameters ${{{\mathbf{w}}}}$ that correspond to the weights and biases of the network. In order to turn this idea into a concrete algorithm, the only missing step is to show how the minimization in eq.~(\ref{tNP}) can be transformed into a neural network training problem. This step is taken in section~\ref{sec:alg}, while here we further elaborate on the conceptual foundations of our method and on the comparison with existing approaches. A brief introduction to neural networks is reported in appendix~\ref{app:NN}.

\subsection{Model-Dependent Tests}\label{sec:MDT}

The Neyman--Pearson formula in eq.~(\ref{tNP}) makes clear that the problem of searching for departures from the reference model expectations (i.e., for new physics) merely reduces to the one of selecting an appropriate alternative hypothesis. Different choices produce different test statistics, with widely different performances. One extreme situation is when compelling theoretical arguments allow us to select a single (simple) alternative hypothesis ``NP'', with no free parameters, for how new physics should look like. In this case eq.~(\ref{tNP}) reduces to
\beq\label{tideal}
\displaystyle
t_{\rm{id}}({\mathcal{D}})=2\,\log\left[\frac{e^{-N({\rm{NP}})}}{e^{-N({\rm{R}})}}\prod\limits_{x\in {\mathcal{D}}}\frac{n(x|{\rm{NP}})}{n(x|{\rm{R}})}\right]\,.
\eeq
According to the so-called Neyman--Pearson lemma \cite{10.2307/91247}, $t_{\rm{id}}$ is the optimal discriminant between the reference and the new physics hypotheses. It is the one that produces the smallest median $p$-value if ${\rm{NP}}$ is the true distribution of the data sample.\footnote{The theorem says that the condition $t_{\rm{id}}>t_c$ defines the critical region with highest power $1-\beta\equiv P(t_{\rm{id}}>t_c|{\rm{NP}})$ at given size $\alpha\equiv P(t_{\rm{id}}>t_c|{\rm{R}})$ \cite{Patrignani:2016xqp}. This statement coincides with the one above because $1-\beta$ is a monotonically increasing function of $\alpha$ and the median $p$-value is the value of $\alpha$ that corresponds to $\beta=1/2$.} We denote this test statistic as ``ideal'' because it is the one which is most suited to discover data departures from the reference model, but we can use it only when the true data distribution is known a priori. 

In the following we employ the ideal test statistic as a figure of merit to assess the performances of our method. However apart from this it is clear that it cannot play a role in the design of model-independent new physics searches, where the goal is to be as agnostic as possible on the alternative hypothesis. Notice indeed that any unjustified assumption on the alternative hypothesis can result in complete loss of sensitivity. For instance suppose that an ideal test is constructed by taking NP to be a narrow resonant peak in an invariant mass distribution, on top of a smoothly falling SM background. The distribution ratio $n(x|{\rm{NP}})/n(x|{\rm{R}})$ appearing in eq.~(\ref{tideal}) is nearly equal to $1$ (hence its log is zero) in the whole mass range, aside from a narrow region around the resonance mass where it is larger. Therefore only the events that fall in that region  contribute to $t$. This is perfectly fine if the resonance is present in the data just as we predicted it, because in this case signal events will fall in that region producing a large $t$ and in turn a small $p$-value. However if the resonance mass is different from the one we assumed, signal events will fall outside that region and they will not contribute to $t$. Therefore even if the resonance truly exists in the data the ideal test would completely miss it.

Several ways exist to mitigate the model-dependence of the ideal test, still remaining within the domain of ``partially model-dependent'' new physics searches. For instance the {\sc{BumpHunter}} \cite{Choudalakis:2011qn} approach essentially employs a composite alternative hypothesis with $3$ free parameters that correspond to the resonance production rate, width and mass. The maximum likelihood fit to the parameters gives a $n(x|{{\mathbf{\widehat{w}}}})$ distribution which resembles the one of the true peak, making signal events automatically fall in the region where $n(x|{{\mathbf{\widehat{w}}}})/n(x|{\rm{R}})$ is large such that their contribution to $t$ is large. This method ensures good sensitivity to a generic resonance, but of course it is completely blind to signals that are non-resonant, or that display a resonant peak in a different kinematical variable than the one that has been selected for the test. More generally one can construct tests based on signal topologies, by assuming the production of a certain type of particle (or particles) with certain decay chains, and modeling the production and the decay in terms of phenomenological parameters. 

\subsection{Model-Independence and Neural Networks}\label{sec:NNMI}

We call ``model-independent'' a new physics search where the alternative hypothesis does not follow from physical considerations, but rather it is selected for technical convenience, with the aim of defining a test that is sensitive to the largest possible variety of putative signals. We have seen that being able to mimic the true underlying distribution is essential for a successful test. Therefore flexibility, i.e. the ability to approximate many functions, is the first important requirement on the set of functions ${\mathcal{F}}$ that define the alternative distribution through eq.~(\ref{alth}). Piecewise constant functions are the most standard and widely employed approximants. Hence it is not surprising that this choice of ${\mathcal{F}}$ produces the binned histogram goodness-of-fit test\footnote{As the name suggests, this test is typically discussed (see e.g.~\cite{Cowan:1998ji}) in the context of parameters fitting, where the histogram is employed to fit a number ``$m$'' of parameters that characterize the expected distribution.} 
, which is the simplest approach to model-independent new physics searches. This test is constructed by dividing the space of observations in bins, and taking ${\mathcal{F}}$ to assume a constant value $w_\alpha$ in each bin $\alpha=1,\ldots N_{\rm{bin}}$. Since each $w_\alpha$ is an independent parameter, the minimization in eq.~(\ref{tNP}) can be trivially performed analytically, giving
\beq\label{eq:gof}
t_{\rm{gof}}({\mathcal{D}})=2\,\sum\limits_{\alpha=1}^{N_{\rm{bin}}}\left[N_\alpha({\rm{R}})-O_\alpha+
O_\alpha\log{\frac{O_\alpha}{N_\alpha({\rm{R}})}}\right]\,,
\eeq
where $O_\alpha$ is the number of counts observed in each bin and $N_\alpha({\rm{R}})$ denotes the expected number in the reference model hypothesis. 

The binned histogram method suffers from well-known limitations, the first one being the arbitrariness in the choice of the binning. A reasonable prescription is to employ the smallest bin size compatible with the experimental resolution on the variable of interest. The second and more severe limitation is that the reach of the goodness-of-fit method is reduced by histogram bins that are in good agreement with the reference model. This point is conveniently illustrated by taking the limit where the number of countings is large in each bin, such that $O_\alpha$ are gaussian-distributed and eq.~(\ref{eq:gof}) reduces to the $\chi^2$ formula. Non-discrepant bins are those where the true model coincides with the reference one, therefore their total contribution to $t$ follows the distribution that is expected in the reference model. A $\chi^2$ with a number of degrees of freedom equal to the number of non-discrepant bins. The mean and the variance of the non-discrepant contribution are thus equal to the number of non-discrepant bins. Instead each bin where there is a discrepancy obviously contributes on average more than a non-discrepant bin, however if there are only a few of them their total contribution can be much smaller than the one of the non-discrepant bins and not appreciably change the total value of $t$. 

Removing non-discrepant bins improves the sensitivity of the test. Hence the binned histogram goodness-of-fit method only works if applied to a restricted set of bins, i.e. to restricted signal regions that have been selected on the basis of prior expectations on the putative signal. Needless to say, the test looses any sensitivity if these expectations are not met by the actual signal. 

As mentioned in the introduction, the problem of non-discrepant bins is not at all an academic one. Existing constraints on new physics models tell us that the vast majority of the data collected in present and future high energy physics and cosmology experiments will agree with the reference model (i.e., the SM and $\Lambda$CDM, respectively). Still we are unable to identify sharply and systematically the data where new physics cannot be present, so ideally the whole set of data will have to be employed in the analysis. This will produce enough non-discrepant bins to wash out essentially any signal that we might expect. Nonetheless the limitations of the binned histogram method can be partially amended, usually at the price of introducing some amount of model-dependence. Approaches based on binned histograms include SLEUTH at D0~\cite{Abbott:2000fb, Abbott:2000gx}, searches at H1~\cite{Aktas:2004pz, Aaron:2008aa}, the VISTA and SLEUTH algorithms at CDF~\cite{Aaltonen:2007dg, Aaltonen:2008vt}, the CMS algorithm MUSiC~\cite{CMS:2008gya, CMS:2011fra}, ATLAS general searches~\cite{ATLAS:2017irs, ATLAS:2012qna, ATLAS:2014sxa}  and \cite{Asadi:2017qon}. 

Here however we want to explore a different direction by questioning the starting point of the construction, i.e. the choice of ${\mathcal{F}}$ as piecewise constant functions. We instead define ${\mathcal{F}}$ as an artificial neural network. It is quite easy to argue against piecewise constant functions and in favor of neural networks and we are not the first ones to do it~\cite{bishop:2006:PRML,Goodfellow-et-al-2016,Haykin1998}. Neural networks are often introduced exactly as a convenient alternative to binned histograms for the estimation of distributions.\footnote{We thank G.~Cowan for explaining this so clearly in his lecture \cite{CowanLecture}.} 

The first argument is that piecewise constant functions are discontinuous and rapidly oscillating. The best fit to the data
\beq\label{f_gof}
f(x;{\widehat{{\mathbf{w}}}})=\left\{\log\frac{O_\alpha}{N_\alpha({\rm{R}})}\;\;{\rm{if}}\;x\in {\rm{bin}}_\alpha,\;{\rm{for}}\;\alpha=1,\ldots,N_{\rm{bin}}\right\}\,,
\eeq
can have large gradients, which randomly assume positive and negative values in adjacent bins, because of statistical fluctuations. Functions of this sort are not at all credible hypotheses on how the true distribution really looks like. Nevertheless these are the ones that we compare with the reference model when we carry out the goodness-of-fit test. Neural networks are on the contrary smooth functions. 

The second advantage of neural networks is that they are more ``efficient'' approximants. Consider a peak of width $\sigma\ll1$ in the distribution of a one-dimensional variable. Reproducing this feature requires a number of bins, i.e. of free parameters, of order $1/\sigma\gg1$ \footnote{A similar estimate applies if we take ${\mathcal{F}}$ to be the Fourier series. Extending the series up to frequencies of order $1/\sigma\gg1$ is needed to see the peak.}. A neural network can instead reproduce (see for instance appendix~\ref{app:NN} and~\cite{NNpedagogical} for a pedagogical introduction) an arbitrarily sharp peak with only $3$ neurons, i.e. with a limited number of parameters. 

Last, but not least, there is the problem of the curse of dimensionality. The number of events that are needed to approximate a function by means of an histogram grows exponentially with the dimensionality of the variable $x$. While a complete proof is still missing, evidence suggests (see for instance \cite{DBLP:journals/corr/PoggioMRML16,DBLP:journals/corr/Bach14,2017arXiv171208688M}) that neural networks can break the curse of dimensionality, requiring fewer events to approximate multivariate distributions. This is of course an extremely desirable property because we would like to search for new physics employing as many variables as possible, reducing in this way the risk of loosing sensitivity because of an erroneous choice of observables. On the other hand we have at our disposal a limited number of events to train the neural network.

\section{The Algorithm}\label{sec:alg}

\begin{figure}[!t]
\centering
\includegraphics[width=1\textwidth]{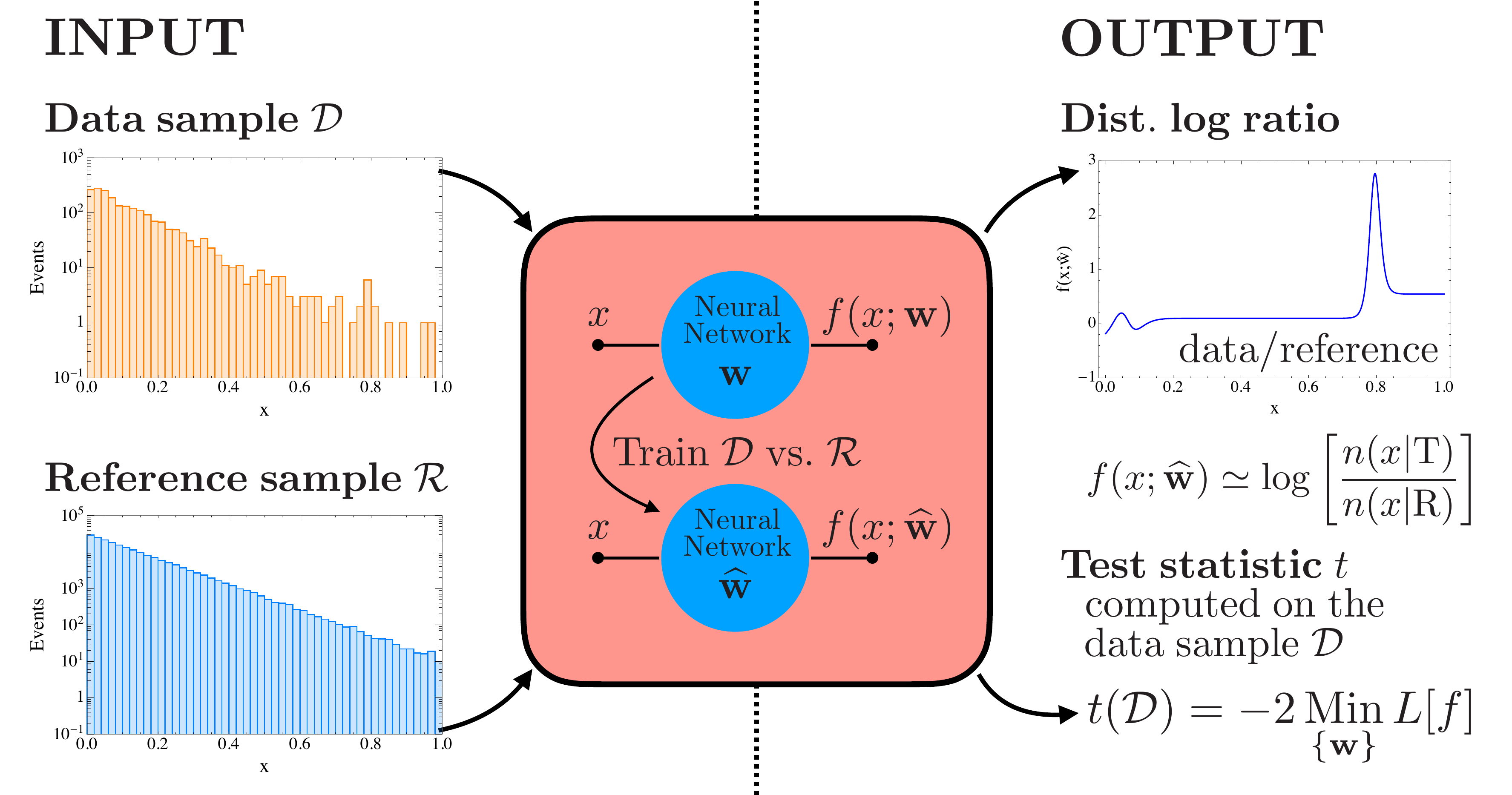}
 \caption{A schematic representation of the implementation of our strategy.}
\label{fig:network_diagram}
\end{figure}

The algorithm aims at comparing a given data sample ${\mathcal{D}}=\{x_i\}$, $i=1,\ldots,{\mathcal{N}}_{{\mathcal{D}}}$,  with the reference model prediction for the distribution of $x$, $n(x|{\rm{R}})$. Normally the prediction does not come in analytical form, but rather in the form of a reference sample ${\mathcal{R}}=\{x_i\}$, with $i=1,\ldots,{\mathcal{N}}_{\mathcal{R}}$, which is distributed according to the reference model. One data and one reference sample are thus the inputs of our algorithm, which produces as output the test statistic $t({\mathcal{D}})$ in eq.~(\ref{tNP}) and the best-fit log-ratio $f(x;{{{\mathbf{\widehat{w}}}}})$. The former quantity will eventually be employed to construct the hypothesis test and turned into a $p$-value as explained at the beginning of section~\ref{sec:bi}. The latter function measures the data disagreement with observation locally in phase space. It can thus be employed to select the most discrepant data for further investigation and to perform a number of sanity checks. A schematic representation of the algorithm is shown in figure~\ref{fig:network_diagram}. A summary of the notation introduced in section~\ref{sec:bi} and in the remainder of this section can be found in table~\ref{tab:notation}.

In the construction of the algorithm we make no explicit assumption on how the reference sample is produced, however we do assume that it is quite large, for example ${\mathcal{N}}_{\mathcal{R}}=100\,N({\textrm{R}})$, in order to eliminate its statistical fluctuations. This is not an issue if the reference sample is produced by a first-principles Monte Carlo event generator, but it might become a problem if instead the reference sample is obtained by extrapolation from a control region. In this case the impact of statistical fluctuations in the reference sample, which we ignore in what follows, should be duly taken into account.

\begin{table}[!t]
\caption{Summary of notation.}
\begin{center}
\begin{tabular}{|c|c|}
\hline
& {\bf Distributions}\\ \hline
$n(x|{\rm{R}})$ & Distribution of the variable $x$ in the reference model ${\rm{R}}$ \\ \hline
$n(x|{\rm{R}})$ & Distribution of the variable $x$ in the new physics model ${\rm{NP}}$ \\ \hline
$n(x|{\rm{T}})$ & True distribution of $x$ \\ \hline
$n(x|\widehat {\bf w})$ & Distribution of $x$ estimated by the Neural Network (NN)\\ \hline \hline 
& {\bf Events}\\
 \hline
$N({\rm{R}})$ & Number of expected events in the reference model ${\rm{R}}$ \\ \hline
$N(\widehat {\bf w})$ & Number of events in the data estimated  by the NN\\ \hline \hline 
& {\bf Test Statistic}\\
 \hline
$t({\mathcal{D}})$ & Test statistic computed by the NN on the data sample $\mathcal{D}$ \\ \hline
$t_{\rm{id}}({\mathcal{D}})$ & Ideal test statistic (requires prior knowledge of the signal) \\ \hline
$P(t|R)$ & Probability distribution of the test statistic $t$ in the reference model R \\ \hline
$P(t|{\rm NP})$ & Probability distribution of the test statistic $t$ in the new physics model NP \\ \hline \hline
& {\bf Normalization}\\ \hline
$\int \hspace{-2pt}n(x) dx =N$ & $n(x)$: Events distribution \\ \hline
$\int \hspace{-2pt}P(x) dx =1$ & $P(x)$: Probability distribution \\ \hline
\hline
\end{tabular}
\end{center}
\label{tab:notation}
\end{table}%

Two problems need to be solved in order to evaluate the test statistic in eq.~(\ref{tNP}) with the elements at our disposal. The first one is that $n(x|{\rm{R}})$ is not known in analytical form, hence we don't know how to compute the integral for $N({{\mathbf{{w}}}})$ in eq.~(\ref{totexp}). The second one is that in order to carry out the minimization numerically, exploiting the powerful existing tools for neural network training, we should first express eq.~(\ref{tNP}) as a loss function. However we can solve both problems at the same time. We estimate $N({{\mathbf{{w}}}})$ by the Monte Carlo method, namely we write \footnote{There is an equality in the equation that follows because we assume a large enough reference sample to reduce the Monte Carlo integration error to a negligible level.}
\beq\label{eq:MCint}
\displaystyle
N({{\mathbf{{w}}}})=\frac{N({\textrm{R}})}{{\mathcal{N}}_{\mathcal{R}}}\sum\limits_{x\in{\mathcal{R}}}e^{f(x;{{{\mathbf{w}}}})}\,.
\eeq
Eq.~(\ref{tNP}) thus becomes
\beq\label{tML}
\displaystyle
t({\mathcal{D}})=
-2\,\underset{\{{\mathbf{w}}\}}{\rm{Min}}\left[\frac{N({\textrm{R}})}{{\mathcal{N}}_{\mathcal{R}}}\sum\limits_{x\in{\mathcal{R}}}(e^{f(x;{{{\mathbf{w}}}})}-1)-\sum\limits_{x\in {\mathcal{D}}}f(x;{{{\mathbf{w}}}})
\right]\equiv-2\,\underset{\{{\mathbf{w}}\}}{\rm{Min}}\,L[f(\,\cdot\,,{\mathbf{w}})]
\,,
\eeq
where $L$ has precisely the form of a loss function. It can be written as a single sum over events by introducing a target variable $y$ which is set to $0$ for the events in ${\mathcal{R}}$ and to $1$ and for those in ${\mathcal{D}}$. Explicitly, we have
\beq\label{eq:loss}\displaystyle
L[f]=\sum\limits_{(x,y)}\left[(1-y)\frac{N({\textrm{R}})}{{\mathcal{N}}_{\mathcal{R}}}(e^{f(x)}-1)-y\,f(x)\right]\,.
\eeq
The minimization of $L$ with respect to the neural network parameters ${\mathbf{w}}$ can thus be carried out as a standard supervised training process. The test statistic is simply minus $2$ times the loss at the end of training. The trained neural network, $f(x;{{{\mathbf{\widehat{w}}}}})$, is the maximum likelihood fit to the data and reference distributions log-ratio. It is the best approximant, within the neural network parametrization, of the true underlying data distribution $n(x|\rm{T})$
\beq\label{eq:appr10}
f(x,{{{\mathbf{\widehat{w}}}}})\simeq \log\left[\frac{{{{n}}(x|{\rm{T}})}}{{{{n}}(x|{\rm{R}})}}\right] \,.
\eeq

Notice that training unavoidably requires some sort of regularization because our loss function (\ref{eq:loss}) is unbounded from below, namely it approaches negative infinity if $f$ diverges at some value of $x$ belonging to the ${\mathcal{D}}$ (i.e., $y=1$) class. Notice that the problematic situation occurs only when the divergence in $f$ is sharply localized, such that $f(x)$ stays finite for all $x\in{\mathcal{R}}$. Otherwise the positive exponent that we have in the loss function for the ${\mathcal{R}}$ (i.e., $y=0$) class overcompensates the negative divergence. We avoid these dangerous configurations by enforcing an upper bound (set by the so-called ``weight clipping'' parameter $W$) on the absolute value of each weight. This forbids the neural network to diverge and to produce sharp features on a scale $\Delta{x}\lesssim1/W$. Given that infinitely sharp features cannot show up in the true distribution because of experimental resolution smearing, for any concrete problem it will be possible to choose $W$ large enough not to limit the approximation capabilities of the neural network. We use $W=100$ in the following.

To obtain a $p$-value that tests the agreement between data and the reference model we proceed as discussed at the beginning of section~\ref{sec:bi}. First we train the network using the actual data sample and a large reference sample distributed according to the R model, as pictorially shown in figure~\ref{fig:network_diagram}. This gives us the observed value of the test statistic $t_{\rm obs}$. Then we repeat the training on many toy experiments generated according to the reference distribution, i.e. we use the same reference sample, network architecture and training parameters as before, but we substitute the data sample with toy reference samples. For each of these samples we compute $t$ and thus obtain $P(t|R)$. The $p$-value is then computed in the usual way (see eq.~(\ref{pv})).

Before moving forward it is worth to clarify some assumptions that our method relies on. First, we assumed knowledge of the expected number of events,  $N({\rm{R}})$, which appears in the definition of the loss function in eq.~(\ref{eq:loss}). This can be problematic because the total event rate is often not well predicted by high energy physics simulations. The simplest way out is to take $N({\rm{R}})$ equal to the number of data that has been observed in the actual experiment. This is conservative as it assumes perfect agreement of the observed number of events with the reference model prediction. In what follows we keep working under the assumption that $N({\rm{R}})$ is known a priori, but this assumption can be easily eliminated as previously explained. Furthermore in real-life applications (and in most of the examples we discuss) the signal component is small and the total number of events is not a significant discriminant. 

Much more problematic is assuming the Monte Carlo to provide a perfect description of the reference distribution shape. This is not realistic because Monte Carlo generators are subject to systematic uncertainties, which for large enough statistics unavoidably result in a significant tension with the data. These uncertainties are routinely modeled as nuisance parameters and treated with the profile likelihood ratio formalism \cite{Rolke:2004mj,Cowan:2010js}. The basic idea is that we should first of all identify the value of the nuisance parameters that best describe the data, taking of course also into account auxiliary measurements and not only the data set of interest. Next we use these values in the reference distribution prediction of eq.~(\ref{tNP}). A proper tune of the reference model Monte Carlo to the data is a prerequisite for any new physics search, hence this problem is in some sense orthogonal to the one that we are addressing. However the interplay and the possible synergies between the two aspects should be carefully studied. Especially the possibility of incorporating in the network the fit to data of some of the nuisance parameters to reduce systematic uncertainties. This is left to future work. 

\subsection{Summary of the algorithm}
\begin{enumerate}
\item Train the network on the data, using the loss function in eq.~(\ref{eq:loss}).
\begin{itemize}
\item {\bf Input}: One data sample $\mathcal{D}$ and one reference sample ${\mathcal{R}}$.
\item {\bf Output}: 1) Value of the test statistic on the data sample $t_{\rm obs}$ and 2) log-ratio of the data and reference probability distribution functions $f(x; \widehat {\bf w}) \simeq \log[n(x|{\rm{T}})/n(x|{\rm{R}})]$.
\end{itemize}
\item Generate several toy data samples ``${\mathcal{D}}$'' that mimic the expected outcome of the experiment if the reference model is true. Train the {\it same} network on these toy data samples, using all the same parameters for training.
\begin{itemize}
\item {\bf Input}: The same reference sample as above and the toy data samples.
\item {\bf Output}: Distribution of the test statistic in the reference hypothesis $P(t|{\rm{R}})$. See for example figure~\ref{fig2}.\end{itemize}
\item Use $P(t|{\rm{R}})$ and $t_{\rm obs}$ to compute the $p$-value: $p=\int_{t_{\rm obs}}^{\infty}P(t|{\rm{R}})dt$. See for example figures~\ref{fig3} and~\ref{fig:NP23} where the $p$-values are reported as $Z$-scores. In those figures we plot a whole set of $p$'s obtained on hundreds of different data samples to assess the performance of our algorithm.
\item If $p$ is sufficiently small to signal a tension with the reference hypothesis, use the log-ratio $f(x; \widehat {\bf w})$ to learn the nature of the discrepancy.
\end{enumerate}

\begin{figure}[!t]
\centering
\includegraphics[width=0.31\textwidth]{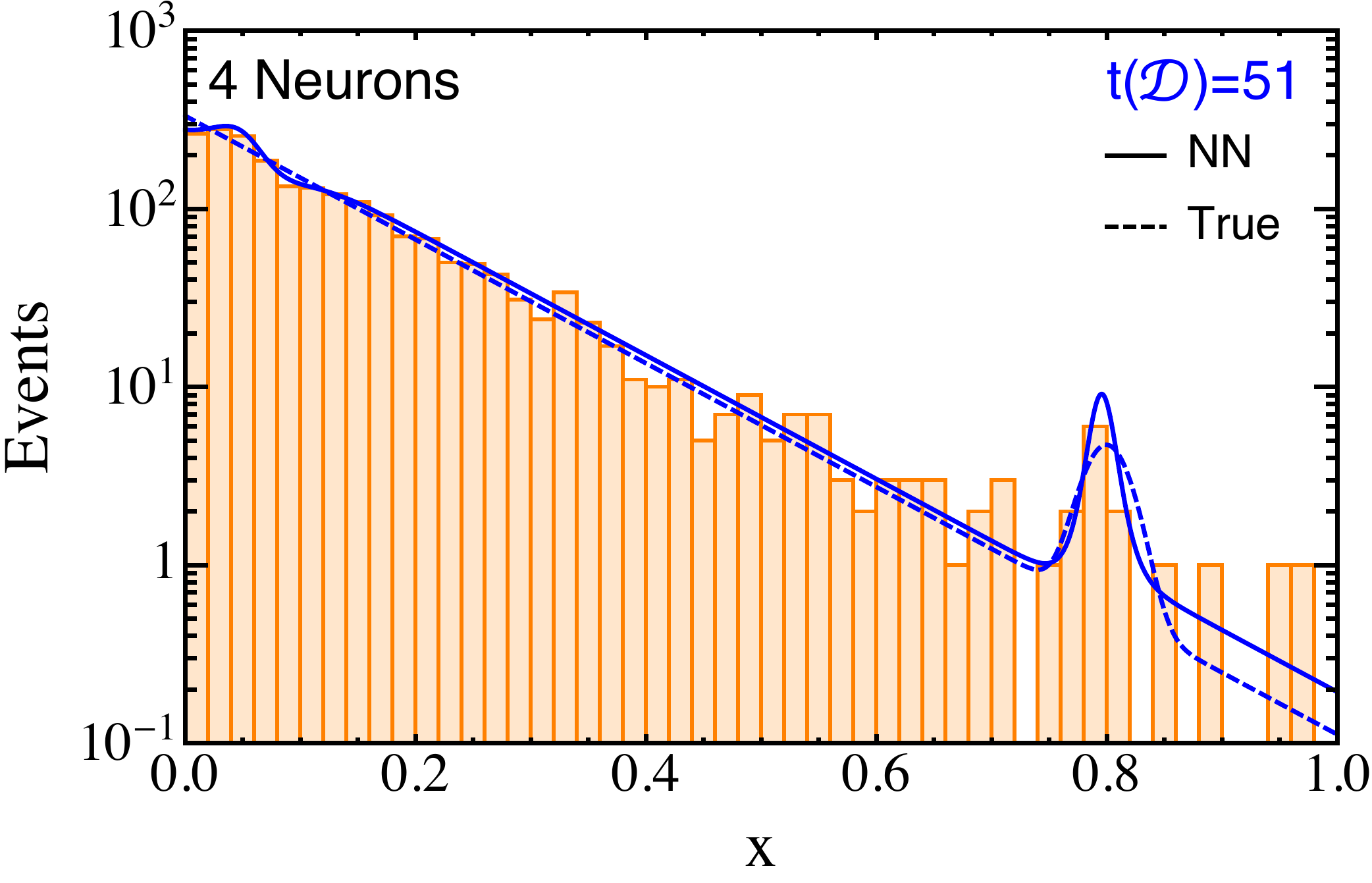}\hspace{10pt}
\includegraphics[width=0.31\textwidth]{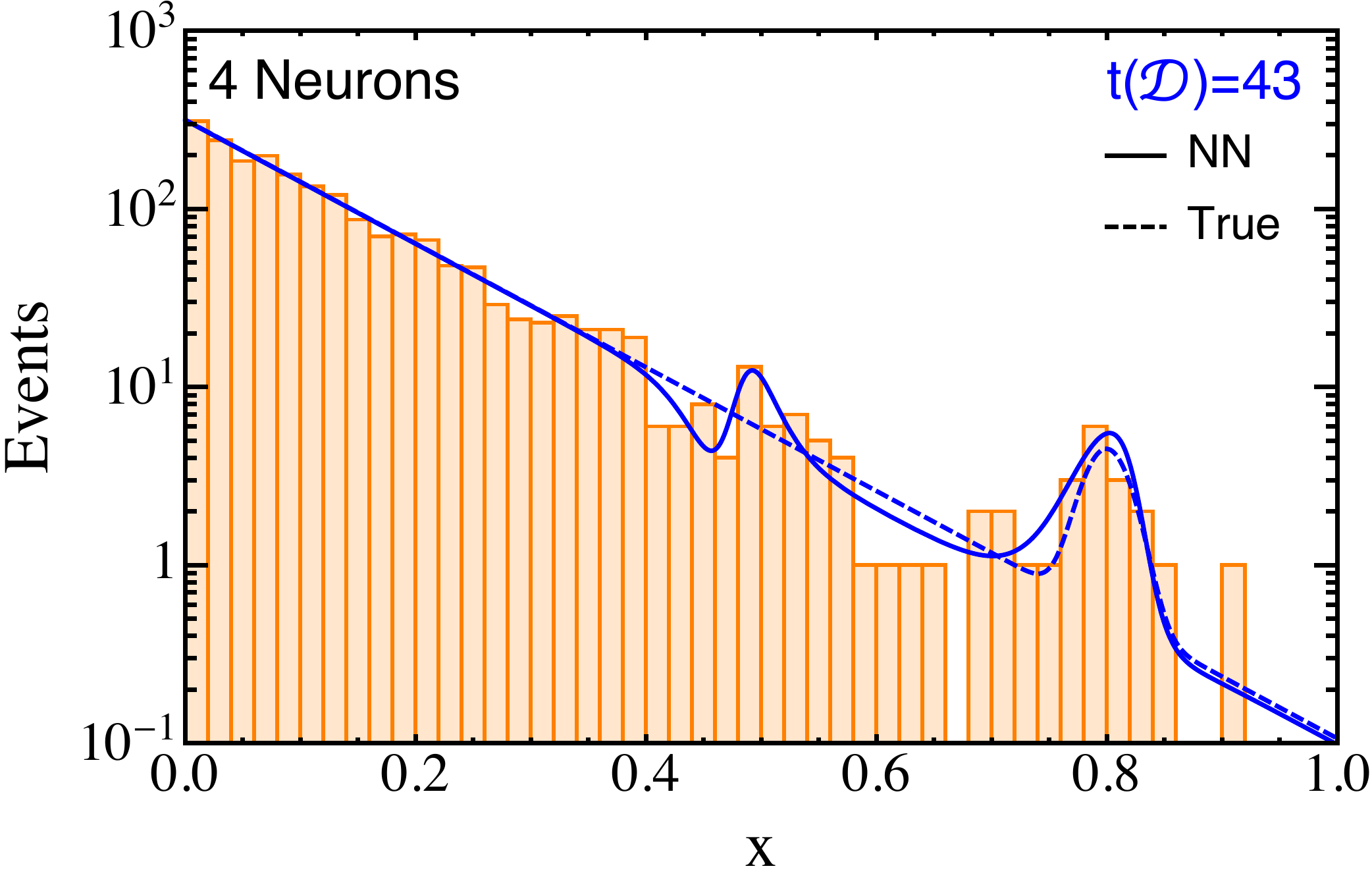}\hspace{10pt}
\includegraphics[width=0.31\textwidth]{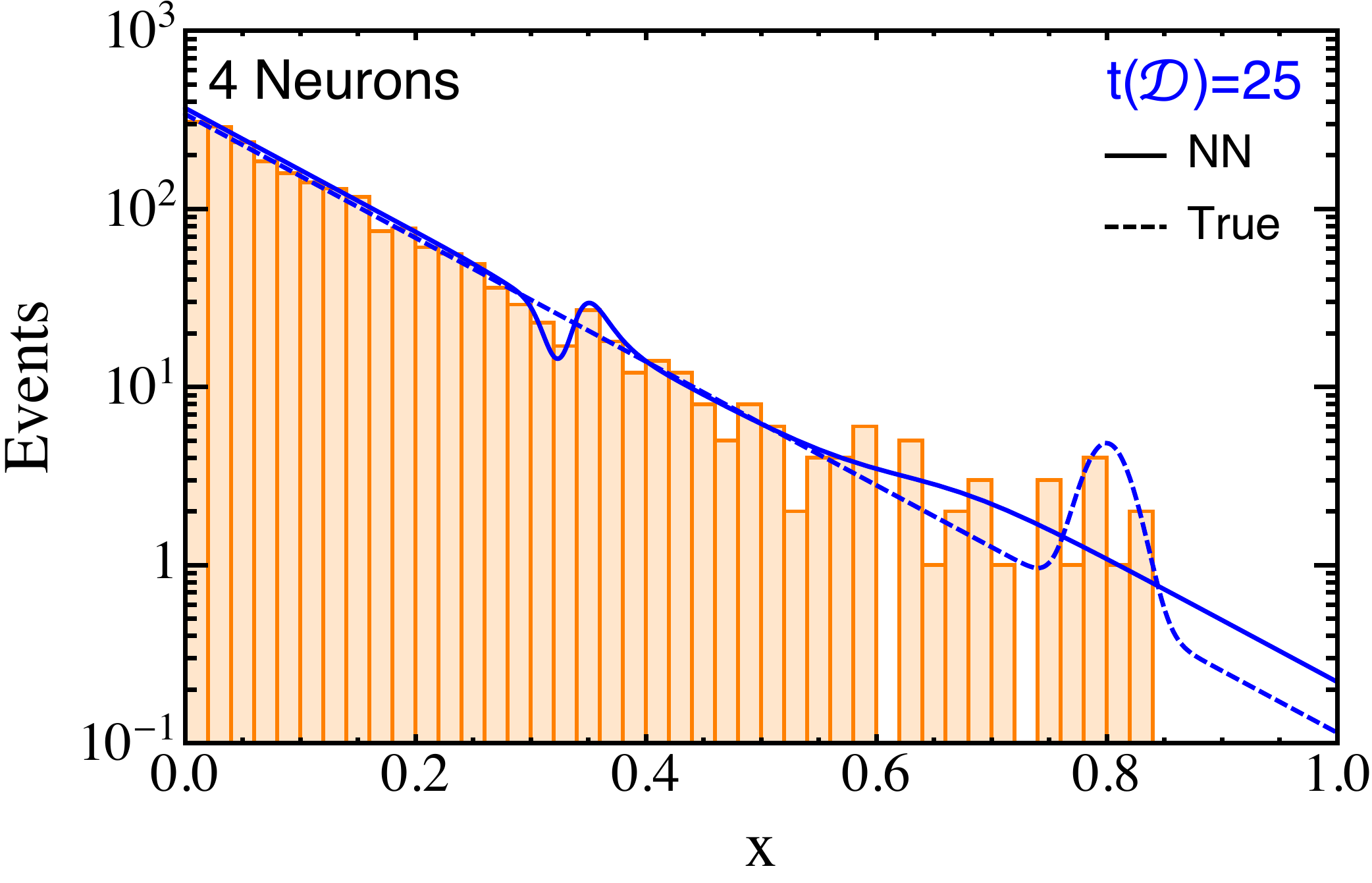}\vspace{10pt}
\includegraphics[width=0.31\textwidth]{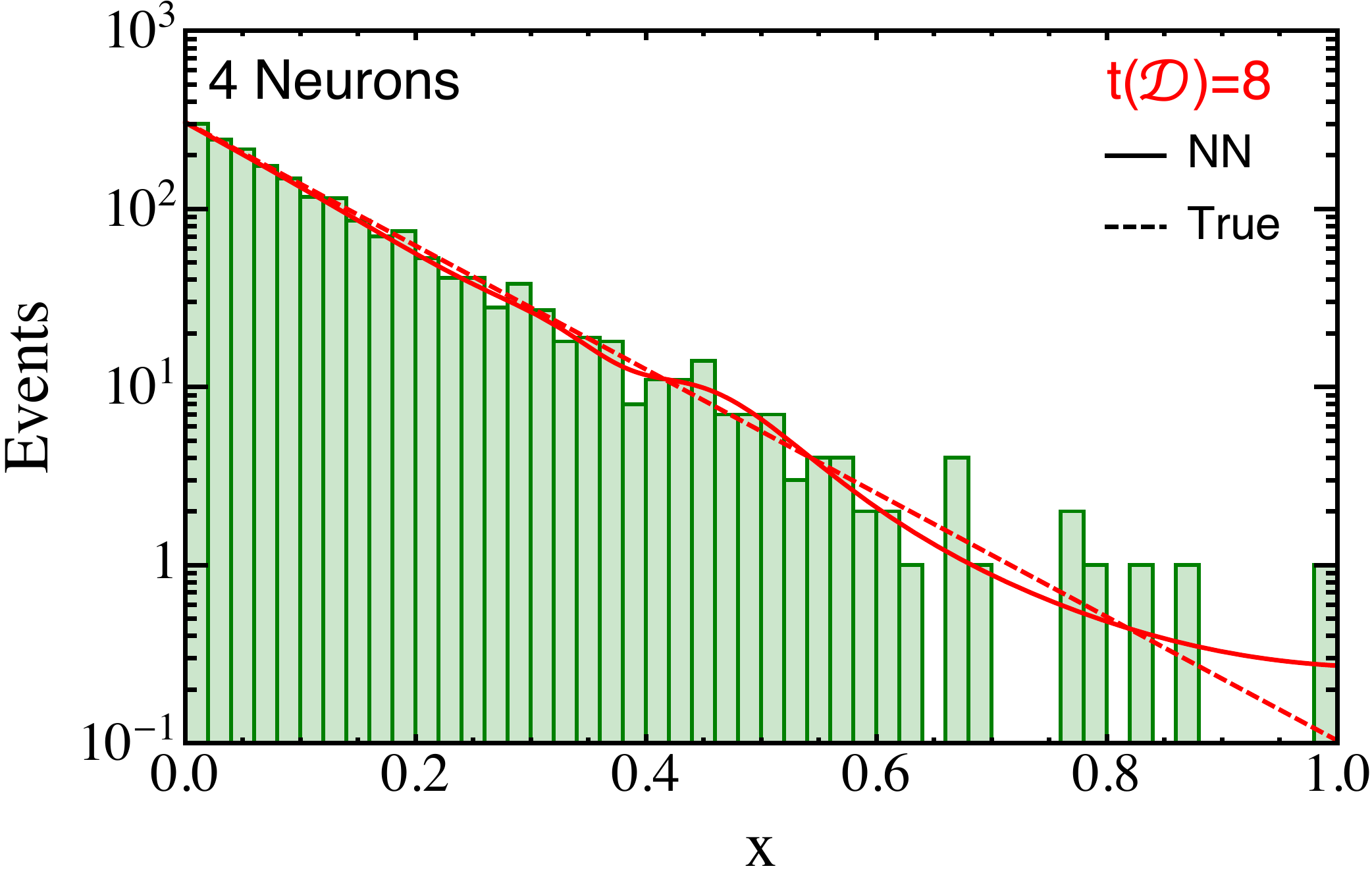}\hspace{10pt}
\includegraphics[width=0.31\textwidth]{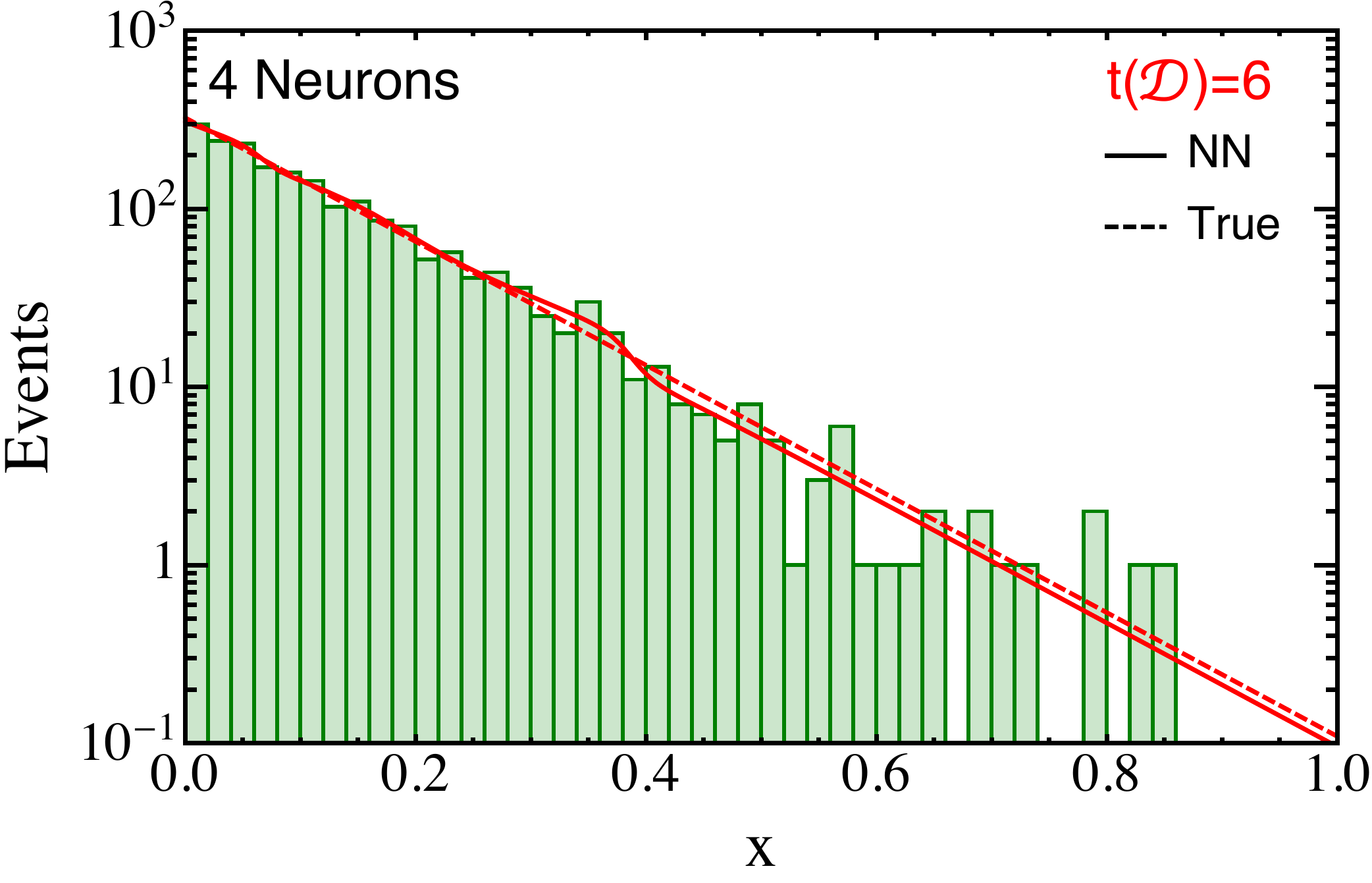}\hspace{10pt}
\includegraphics[width=0.31\textwidth]{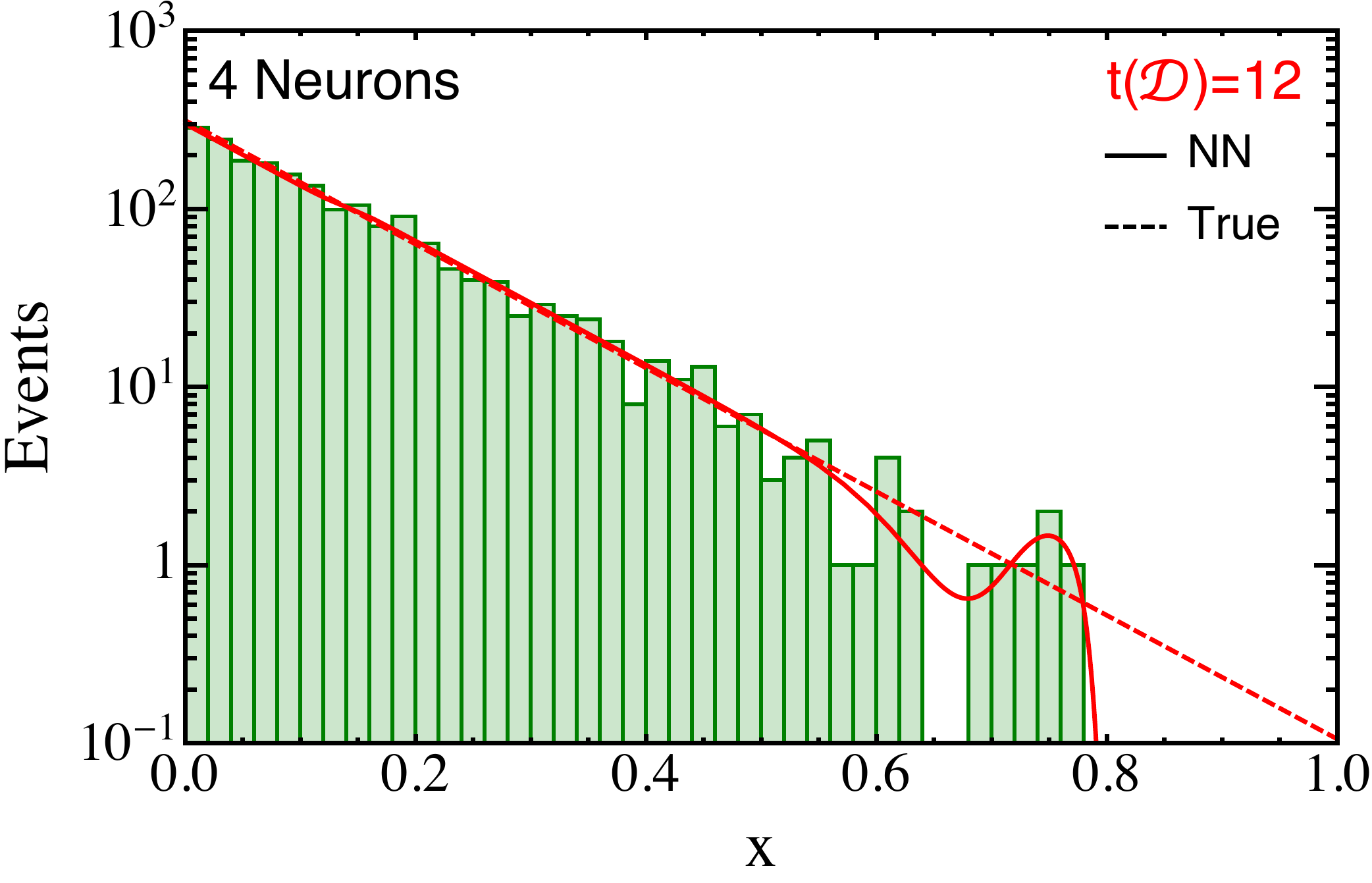}
 \caption{The distribution learned by a neural network with a single 4-neurons hidden layer (solid line), compared with the distribution used to generate the data (dashed line) and the binned histogram of the training data set. The value of the test statistic $t(\mathcal{D})$ obtained by the network is reported in the upper right corner of each plot. The higher values of $t(\mathcal{D})$ in blue signal that the network is discriminating between data sets containing new physics (top row) and data sets following the reference hypothesis (bottom row).}
\label{fig1}
\end{figure}

\subsection{Performances on a Simple Case Study}

We now turn to a first illustration of the performances of our algorithm. We start with a simple example, which we study more quantitatively and systematically in the next section. We consider an univariate problem $x\in[0,1]$. The reference model (or background) is a steeply-falling exponential distribution
\beq\label{SMref1}
P(x|{\rm{R}})\propto e^{-8\,x}\hspace{-2pt}\,,\;\;{\textrm{and}}\;\;{{N}}({\rm{R}})=\int_0^1\hspace{-2pt}dx\,n(x|{\rm{R}})=2000\equiv B\,.
\eeq
We consider the possible presence in the data of a small resonant signal component $S=10$, distributed as
\beq\label{BSM1}
{{P}}(x|{\rm{S}})\propto e^{-\frac{(x-\bar{x})^2}{2\sigma^2}}\hspace{-2pt}\,,\;\;{\rm{with}}\;\bar{x}=0.8\;{\rm{and}}\;\sigma=0.02\,.
\eeq
The new physics distribution for $x$ therefore is
\beq\label{BSM1prime}
{{n}}(x|{\rm{NP}})=\frac{S+B}{1+S/B}\bigg[ 
{{P}}(x|{\rm{R}})+\frac{S}{B}\, {\rm{P}}(x|{\rm{S}})
\bigg]
\,,
\eeq
with a signal over background ratio $S/B=5\cdot10^{-3}$ and a total number of expected events $N({\rm{NP}})=S+B=2010$. We generate one large (${\mathcal{N}}_{\mathcal{R}}=200\,000$) reference sample ${\mathcal{R}}$ according to the reference p.d.f., and several data samples ${\mathcal{D}}$ that follow either the reference or the new physics distributions. The number of data events is selected at random taking into account Poisson fluctuations around the expected numbers $N({\rm{R}})=2000$ and $N({\rm{NP}})=2010$. We train a $4$-neurons $(1,4,1)$ neural network\footnote{The notation for the neural network architecture is explained in more detail in appendix~\ref{app:NN}. The $(1,4,1)$ network has one-dimensional input and output and a hidden layer with 4 neurons.} on each data set and we obtain the corresponding $t({\mathcal{D}})$ and $f(x;{{{\mathbf{\widehat{w}}}}})$ as previously described. Since $n(x|{\rm{R}})$ is fully known, in our toy example we can also compute the best-fit distribution $n(x|{{{\mathbf{\widehat{w}}}}})$ using the log-ratio learned by the neural network in eq.~(\ref{alth}). An initial learning rate of $10^{-3}$ is chosen, and training is stopped after $150\,000$ rounds. The results are displayed in fig.~\ref{fig1} for six representative data samples. The ones on the first and on the second row have been obtained from the NP and from the R distributions, respectively.

The figure illustrates a number of interesting points. First of all, we see that in all cases the distribution learned by the neural network is very much correlated with the data sample that was used for training. Still it doesn't follow the data too closely, producing smooth curves that are quite ``credible'' hypotheses on the true underlying distribution. This should be contrasted with the discontinuous piece-wise constant distribution, i.e. the envelope of the histogram, that one would effectively rely on if the same data sets where studied with the binned histogram method. We also see that in the bulk region, i.e. at small $x$, the neural network is able to reproduce very accurately the true distribution, thanks to the large statistic. This is important because mismodeling the bulk would produce a large spurious contribution to $t$, that would obscure the genuine signal in the tail. The NP-generated data samples produce an excess in the tail of the distribution, which is more or less in agreement with the true peak at $x=0.8$, depending on how many events happened to fall in that region. The distributions obtained with the background data samples can also depart considerably from the reference distribution (which is the true one for background samples), however the departures occur in regions were only few events are present and hence they give a limited contribution to $t$. We also remark that the size of $t$ is in clear correspondence with how different the reference distribution and the distribution learned by the neural network are. The six values shown in the figure already indicate that $t$ possesses some discriminating power between the signal and the background. We study this systematically in the following section.

\section{Numerical Experiments}\label{numexp}

In this section we test our method by performing several numerical experiments on one and two-dimensional samples.  A summary of the notation needed to interpret the figures in this section can be found in table~\ref{tab:notation}. In all the new physics scenarios discussed here we have generated hundreds of toy data samples to assess the median significance of the algorithm and its correlation with the ideal significance. So the single value of the test statistic, $t_{\rm obs}$, that one would observe in a real experiment is presented as a distribution given a putative new physics model. Correspondingly the single observed $p$-value (or $Z$-score) becomes an entire distribution.

The numerical experiments performed here have been selected with the aim of illustrating the following aspects:

\begin{itemize}
\item {\emph{Model-Independence:}} The goal of our approach is to be sensitive to a signal that is unknown a priori. Ideally it should detect any kind of new physics that could be present in the data. We verify this through several examples in section~\ref{sec:MI}. 
\item {\emph{(In-)Sensitivity to cuts:}} It is impossible to identify the appropriate search region without prior assumptions on the nature of the  signal. Furthermore we argued in section~\ref{sec:alg} that the loss of sensitivity due to the presence of a large number of data points in agreement with the reference model, is the main limitation of the binned histogram goodness-of-fit approach. In section~\ref{sec:cuts} we show that instead the performances of our method do not depend on whether a favorable signal region is selected based on prior knowledge of the signal.
\item {\emph{Two dimensions:}} We apply our method to two-dimensional distributions, with the aim of studying to what extent the reach deteriorates if the relevant variable that differentiates the signal from the background is not known a priori. The results are presented in section~\ref{2d}.
\item {\emph{Dependence on hyperparameters:}} The neural network architecture, the initial learning rate and the number of training rounds are the free parameters of our algorithm, collectively denoted as hyperparameters. We study the performance dependence on these parameters in section~\ref{hyp}. 
\end{itemize}

Before discussing these points, a general methodological remark is in order. It is not completely straightforward to quantify the performances of our method. Clearly in each example we can compute the median $p$-value of our test, but this is a valid figure of merit only in comparison with some independent quantification of the actual difference between the reference distribution and the new physics that we assumed in the example. This aspect is particularly important for comparing the sensitivity of our test to new physics signals of different nature, for instance comparing the sensitivity to a peak with the one to an anomalous growth of the distribution in the tail. What we need is to assess in absolute terms how difficult it is to discover new physics in the example under consideration. For this purpose we employ the ``ideal'' test, defined in eq.~(\ref{tideal}). Namely for each toy example we evaluate $t_{\rm{id}}$, defined by exploiting the complete knowledge of the new physics distribution, on a large set of reference-distributed toy data samples. This gives us the p.d.f. of $t_{\rm{id}}$ in the reference hypothesis. Next we use this distribution to compute the ideal $p$-value $p_{\rm{id}}$ for each one of the toy data samples generated according to the new physics distribution. The ideal $p$-value can then be compared with the one obtained with our test, either individually on each sample or globally in terms of the median over repeated toys. Notice that the ideal test is the one with smallest median $p$-value, since it is obtained using a complete knowledge of the signal. Therefore we cannot hope to obtain a similar significance with our test, where we assume no previous knowledge of the signal whatsoever. Still we can asses the success or failure of our method by how much significance we lose in comparison with the ideal test.

\subsection{Model-Independence}\label{sec:MI}

\begin{figure}[t]
\centering
\includegraphics[width=0.95\textwidth]{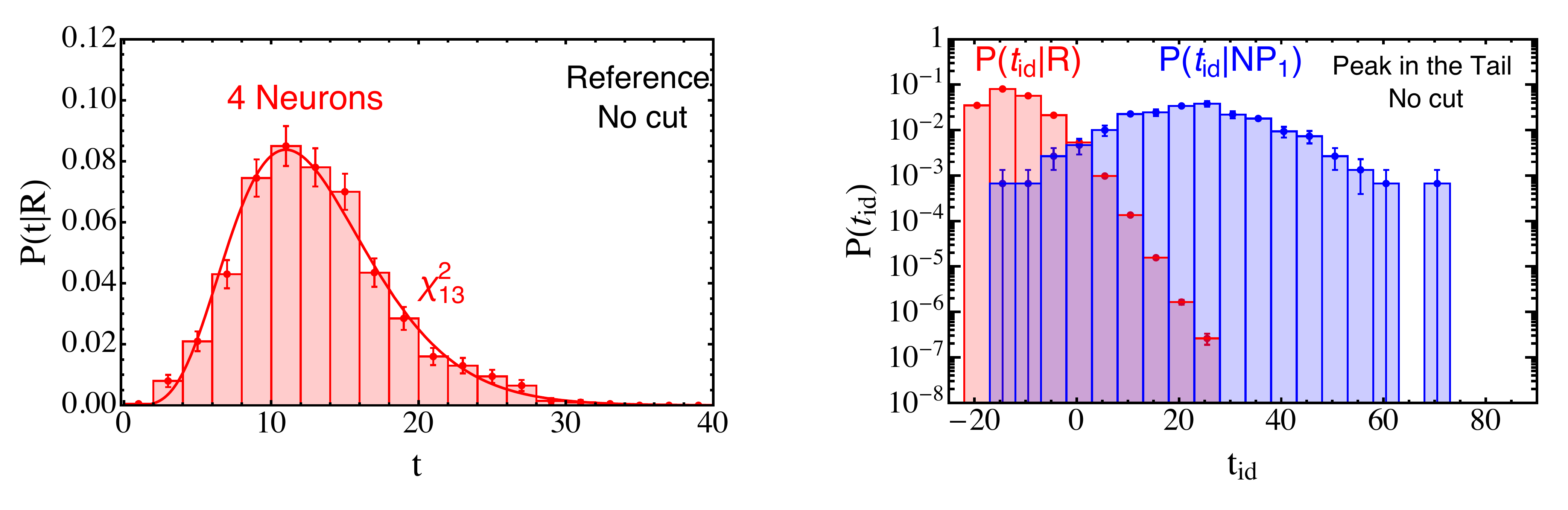}
\caption{{\it Left panel}: Test statistic distribution in the reference model, compared with the $\chi^2$ p.d.f. with $13$ degrees of freedom. The relation between the $\chi^2$ and our test statistic is discussed in sections~\ref{sec:MI} and \ref{hyp}. {\it Right Panel}: Ideal test statistic distribution in the reference and in our first new physics scenario: NP$_1$.}
\label{fig2}
\end{figure}

In all the examples considered in the present subsection, $x\in[0,1]$ and its reference distribution is the exponential in eq.~(\ref{SMref1}). Physically we might interpret $x$ as an invariant mass measured at the LHC, with its steeply-falling SM distribution modeling parton luminosities. Since the reference distribution is the same for all example signals, the preparatory stages of our test can be carried out once and for all. These consist in generating a ${\mathcal{N}}_{\mathcal{R}}=200\,000$ reference sample and in computing the test statistic p.d.f. by training the neural network on toy Monte Carlo samples generated according to the reference model. A $(1,4,1)$ neural network is employed, the initial learning rate is $10^{-3}$ and $150\,000$ training rounds are performed using the \textsc{RMSprop} algorithm~\cite{RMSprop}. Evaluating $t({\mathcal{D}})$ on $1000$ reference-distributed toys produces the p.d.f. in the left panel of figure~\ref{fig2}. Thanks to this distribution we can compute the $p$-value associated with $t({\mathcal{D}})$ evaluated on the data samples generated according to the new physics distribution. 

Notice however that we can meaningfully estimate the $p$-value only if $t$ does not exceed the maximal value obtained with our toy Monte Carlo samples. If $t$ is larger we can only set a lower bound on the $p$-value, which we obtain from the $68\%$  upper limit for $0$ successes (binomially distributed) and $N$ trials, i.e. $p<1-(0.32)^{1/N}$. With the $N=1000$ Monte Carlo samples at our disposal, this corresponds to $p<1.1\,10^{-3}$ or to a significance $Z>3.05\,\sigma$.\footnote{We adopt the standard definition $Z=\Phi^{-1}(1-p)$, where $\Phi^{-1}$ is the quantile of the Gaussian distribution.} However $P(t|{\rm{R}})$ is quite well approximated by a $\chi^2$ distribution with $13$ degrees of freedom, which is not surprising because $13$ is the number of free parameters of the $(1,4,1)$ network that we are employing. We return on this point in section~\ref{hyp}, for the moment we just exploit this fact to extend our estimate of the significance to values of $t$ above the maximum. Namely, for those we report the estimate of the significance obtained with the $\chi^2$ approximation, instead of the lower bound obtained with the toys.

\begin{figure}[t]
\centering
\includegraphics[width=0.95\textwidth]{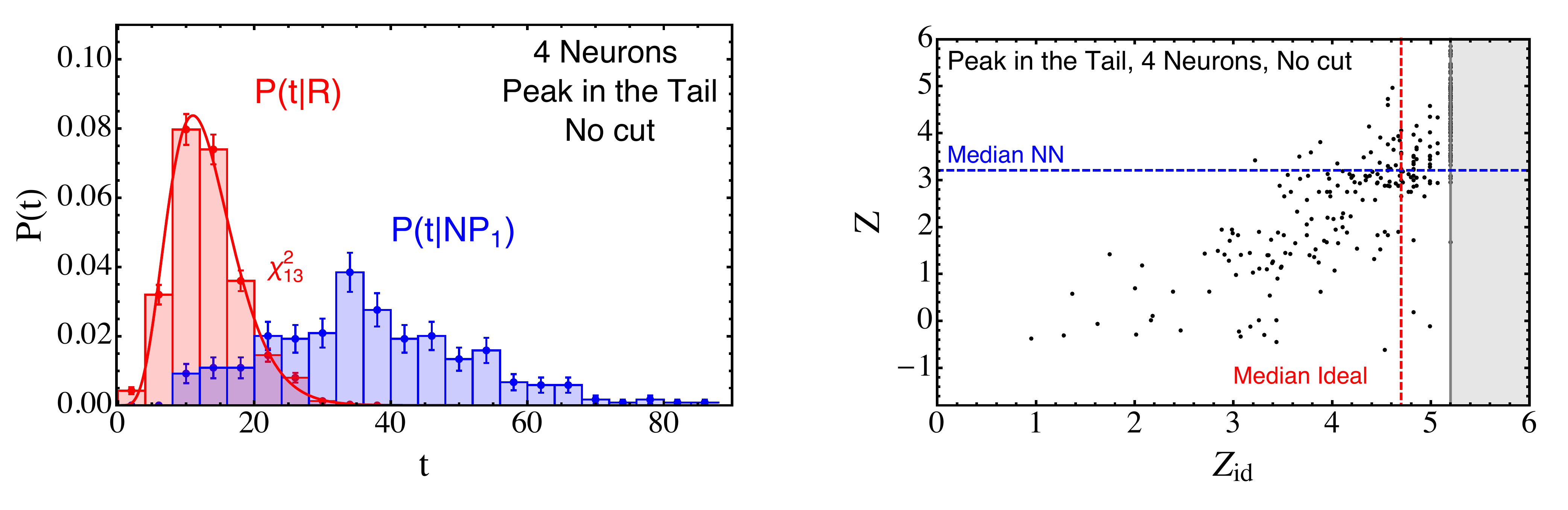}\hspace{15pt}
\caption{{\it Left panel}: Test statistic distribution in the NP$_1$ new physics model $P(t|{\rm{NP}}_1)$, compared with the reference one $P(t|{\rm{R}})$. The two models are defined in equations~(\ref{SMref1}) and (\ref{BSM1}), respectively, and shown in figure~\ref{fig:models}. The larger values of $t$ in $P(t|{\rm NP}_1)$ compared to $P(t|{\rm R})$ signal that our algorithm is sensitive to this new physics scenario. These two distributions are used to obtain the $Z$-score on the y-axis in the right panel. {\it Right panel}: Correlation between the significances (expressed in number of $\sigma$'s) of our test and of the ideal test defined in section~\ref{sec:bi}, for the NP$_1$ model. The gray shaded area corresponds to the region where the ideal significance can not be computed with the number of toy data sets generated. We also show the median significance of our algorithm (Median NN) and the ideal one.}
\label{fig3}
\end{figure}

The first new physics model that we discuss (dubbed NP$_1$ in what follows) is the one introduced in eq.s~(\ref{BSM1}) and (\ref{BSM1prime}). It mimics the presence of a resonance in the tail of the SM invariant mass distribution. We generate $300$ toy Monte Carlo samples according to the new physics distribution in eq.~(\ref{BSM1prime}), and we train a neural network for each, with the same algorithm used for the reference-distributed data. The resulting distribution for $t$, $P(t|{\rm{NP}}_1)$ is displayed in the right panel of figure~\ref{fig3}. By comparing with $P(t|{\rm{R}})$ we see that our test statistic has a considerable discriminating power between the two hypotheses. The median $t$ in the NP$_1$ toy samples is $36$, which is slightly above the maximum value that we obtained with the reference data. The median significance for the NP$_1$ signal hypothesis is thus above $3.05\,\sigma$, and it can be estimated to be $3.2\,\sigma$ using the $\chi^2$ approximation.

\begin{figure}[!t]
\centering
\includegraphics[width=0.31\textwidth]{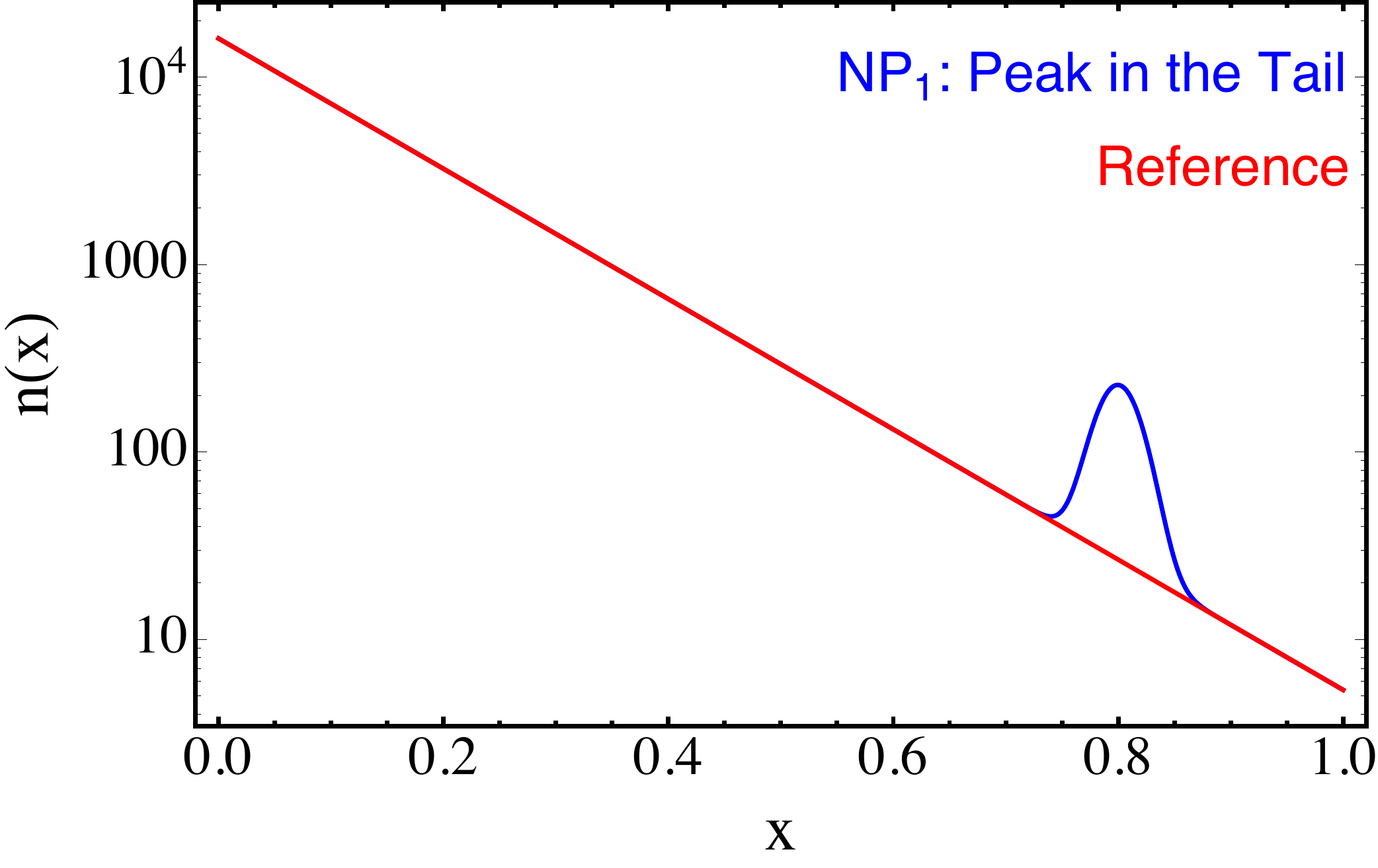}\hspace{10pt}
\includegraphics[width=0.31\textwidth]{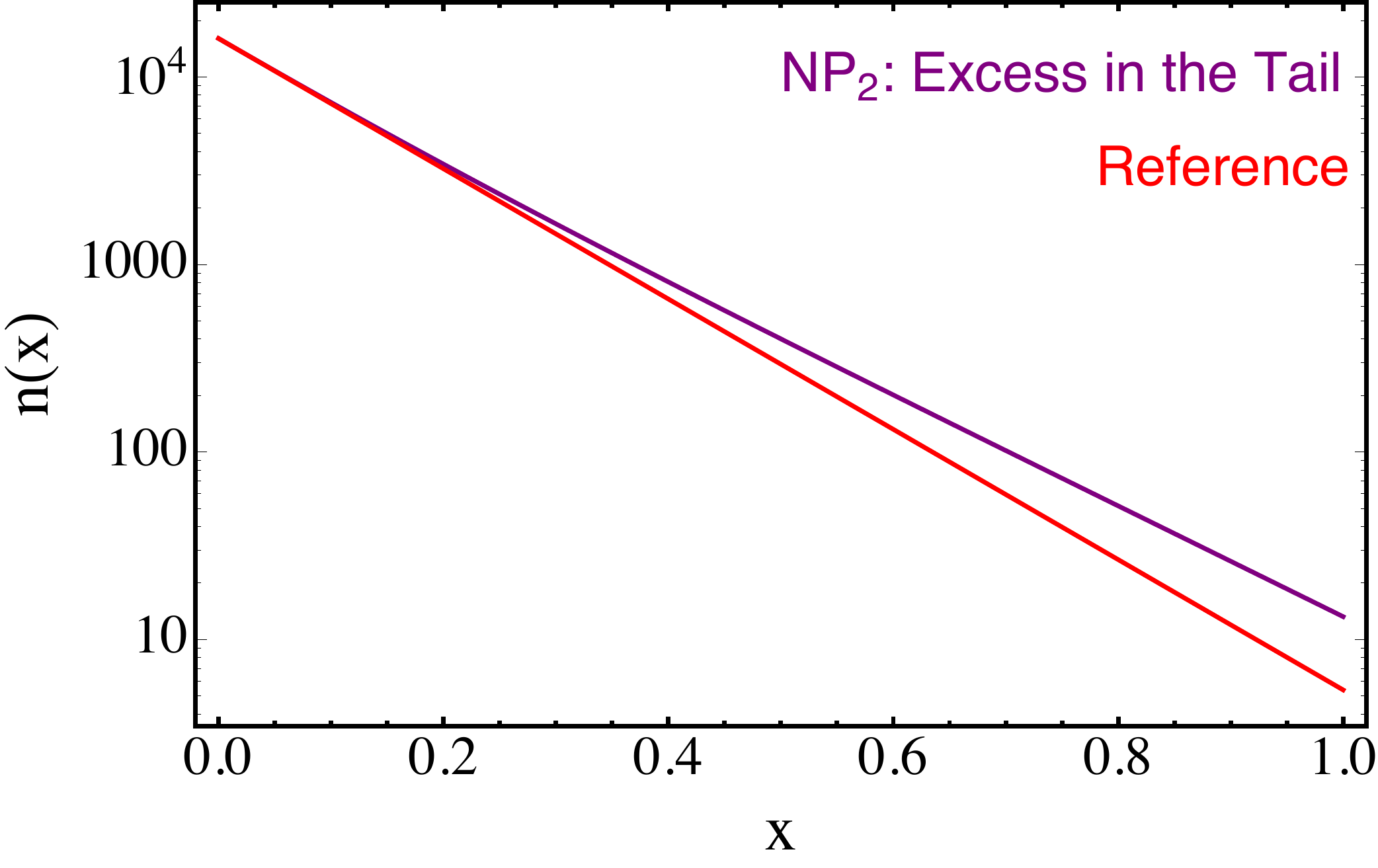}\hspace{10pt}
\includegraphics[width=0.31\textwidth]{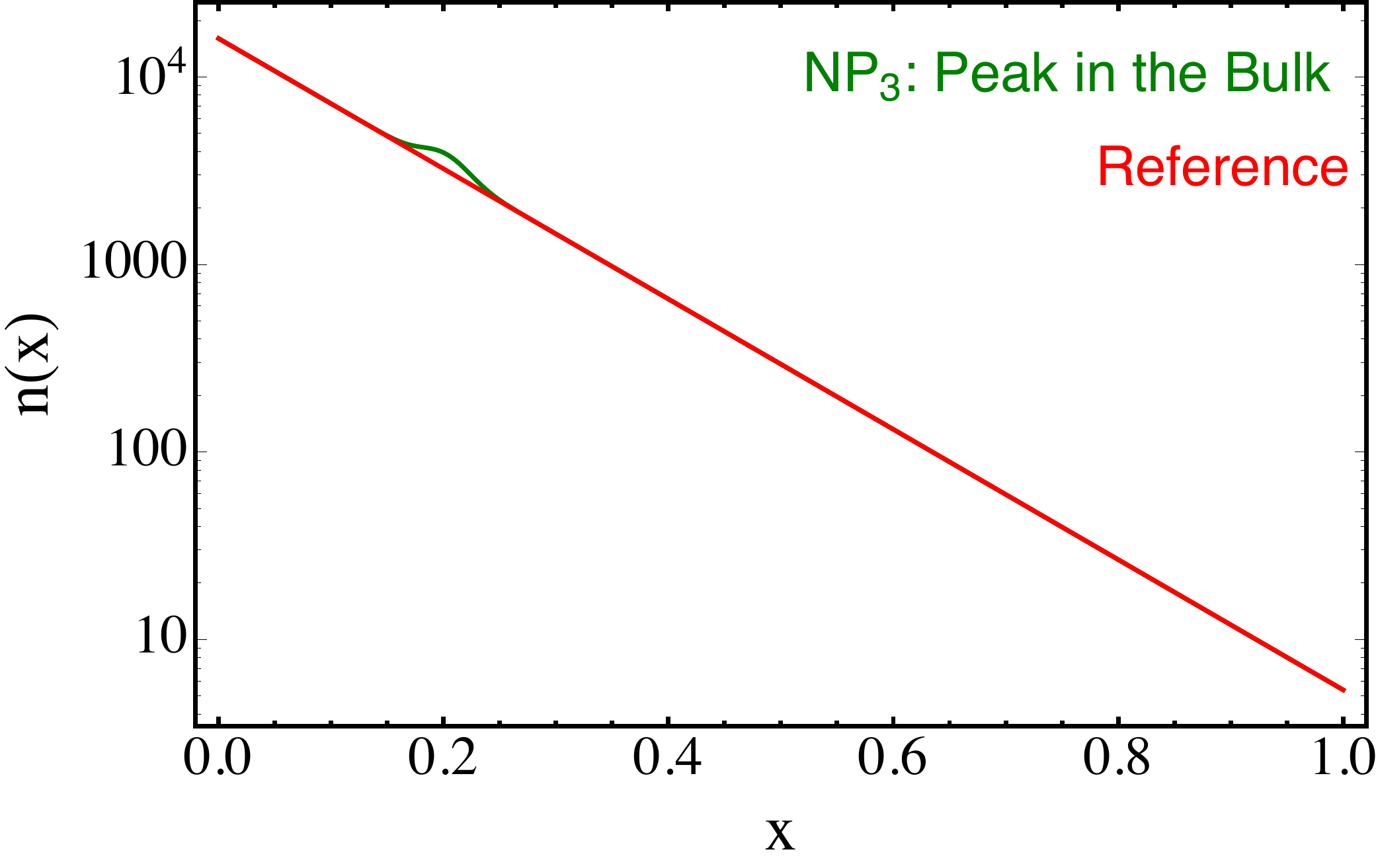}\vspace{10pt}
\caption{The distributions of the three new physics models used in this work plus the reference one. 
}
\label{fig:models}
\end{figure}

For a better assessment of the performances of our method we compare them to those of the ideal test presented in section~\ref{sec:bi} (see the discussion below eq.~(\ref{tideal})). We estimate the ideal test statistic p.d.f. by means of a very large set of $10\,000\,000$ reference model toy data samples, and we compare it with the values of $t_{\rm{id}}$ on the $300$ new physics data samples with which we trained the network. The result is shown in the left panel of figure~\ref{fig2}. The sensitivity of the ideal test is as expected much higher than ours. The median $t_{\rm{id}}$ on new physics samples is $23$ and it corresponds to an ideal significance $Z_{\rm{id}}=4.7\,\sigma$. We can thus conclude that the difference in sensitivity amounts to roughly $1.5\,\sigma$. This is confirmed if we look at the correlation between $Z_{\rm{id}}$ and $Z$ on each individual data sample, reported in the right panel of figure~\ref{fig3}. Notice that the vertical band of points that seemingly breaks the correlation is an artifact due to new physics samples with a $t_{\rm{id}}$ that is larger than the maximum $t_{\rm{id}}$ obtained in the $10\,000\,000$ reference toys. For these samples, a lower bound on $Z_{\rm{id}}$ of $5.2\,\sigma$ (corresponding to zero observed over $10\,000\,000$ trials at  $68\%$~CL) is reported in the plot.

\begin{figure}[!t]
\centering
\includegraphics[width=0.95\textwidth]{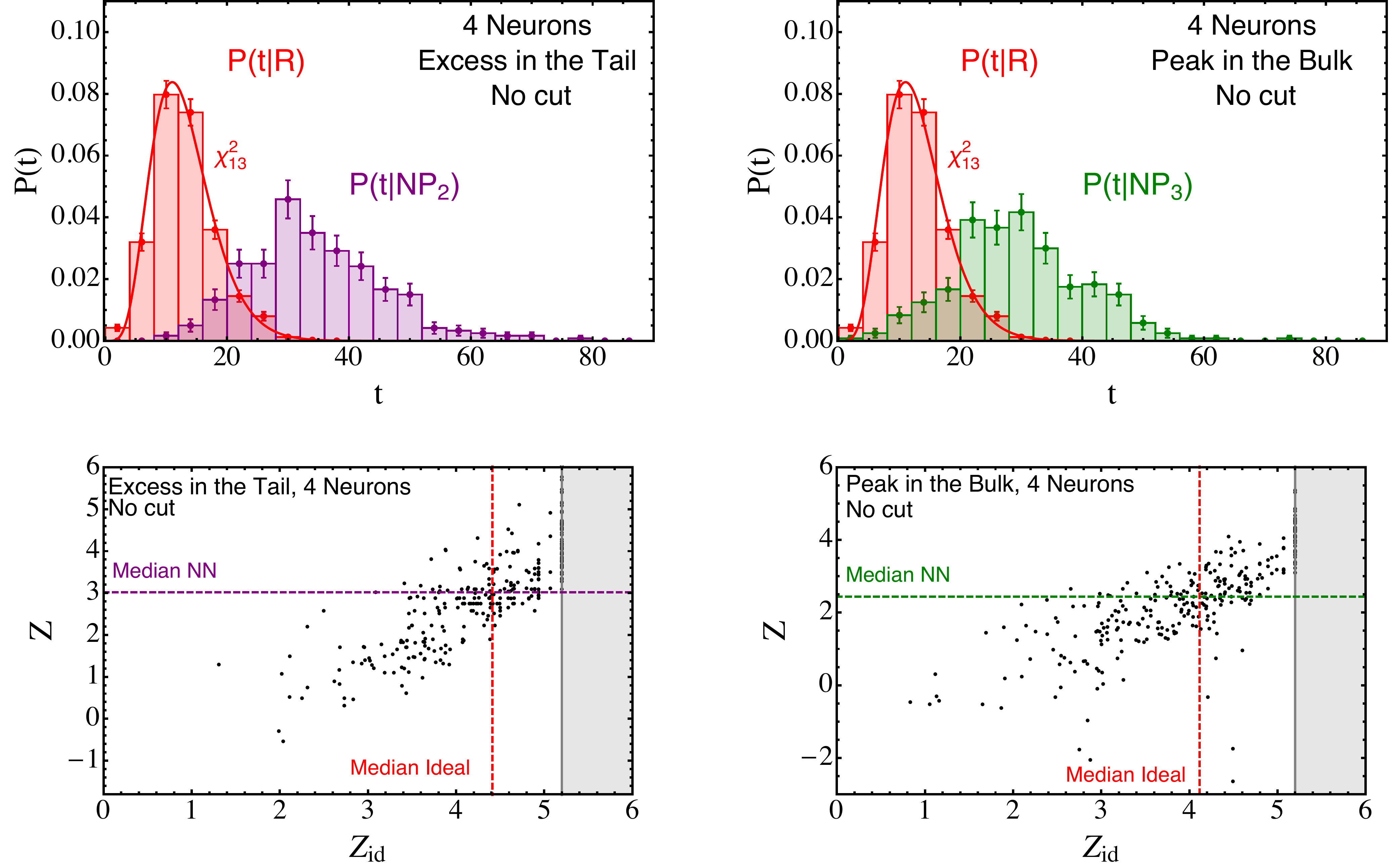}\hspace{15pt}
\caption{{\it Top row}: Test statistic distribution in the NP$_2$ (left) and NP$_3$ (right) new physics models, compared with the reference one.  The two models are defined in eq.~(\ref{BSM2}) and eq.~(\ref{BSM3}). {\it Bottom row}: Correlation between the significances (expressed in number of $\sigma$'s) of our test and of the ideal test defined in section~\ref{sec:bi}, for the NP$_2$ (left column) and NP$_3$ (right column) new physics models. The gray shaded area corresponds to the region where the ideal significance can not be computed with the number of toy data sets generated. We also show the median significance of our algorithm (Median NN) and the ideal one.}
\label{fig:NP23}
\end{figure}

The second example (NP$_2$) is non-resonant new physics, showing up as a quadratic growth with energy in the tail of the reference model distribution. In this case the signal is distributed as
\beq\label{BSM2}
{{P}}(x|{\rm{S}}_2)\propto x^2 e^{-8\,x} \,,
\eeq
and the total expected number of signal event is taken to be $S=90$. Signal and background are combined to define the NP$_2$ distribution as in eq.~(\ref{BSM1prime}). The median ideal significance for the chosen value of $S$ equals $4.4\,\sigma$, very much comparable with the one of the NP$_1$ signal. This ensures a fair comparison between the two. The performances of our algorithm, shown in the left column of figure~\ref{fig:NP23}, are essentially identical to those we obtained for NP$_1$. The median significance is $3.1\,\sigma$ and the correlation between $Z_{\rm{id}}$ and $Z$ again reveals a significance loss of around $1.5\,\sigma$.

Finally, we discuss another resonant signal, emerging this time in the bulk of the reference model distribution. The signal distribution is
\beq\label{BSM3}
{{P}}(x|{\rm{S}}_2)\propto e^{-\frac{(x-\bar{x})^2}{2\sigma^2}}\hspace{-2pt}\,,\;\;{\rm{with}}\;\bar{x}=0.2\,,\;\sigma=0.02\,,
\eeq
and $S=35$. The median ideal significance is $4.1\,\sigma$. We see in the right column of figure~\ref{fig:NP23} that accordingly the median significance of our algorithm (2.6$\,\sigma$) is slightly reduced compared to NP$_1$ and NP$_2$. The correlation between $Z_{\rm{id}}$ and $Z$ is equally sharp. 

The comparative study of three new physics models carried out in this section provides a clear confirmation of the model-independent nature of our approach.

\begin{figure}[!t]
\centering
\includegraphics[width=0.95\textwidth]{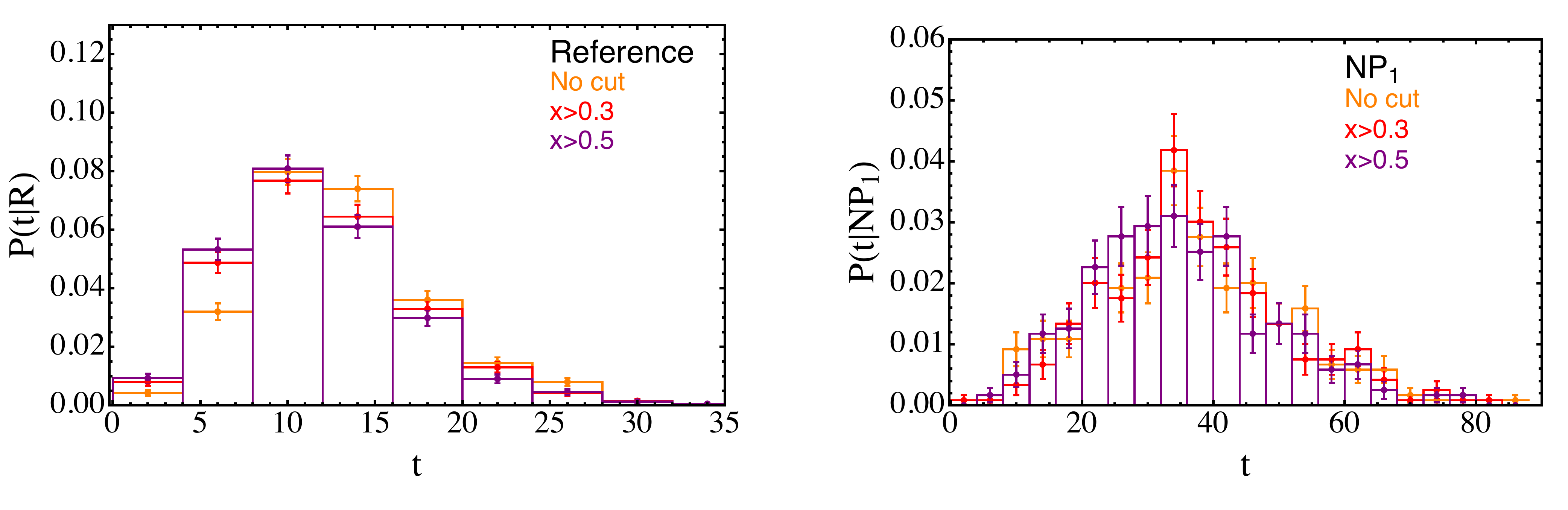}\hspace{15pt}
\caption{Left panel: Test statistic distribution in the reference hypothesis, $P(t|R)$, for $x\geq0, x>0.3$ and $x>0.5$. Right panel: Test statistic distribution in the new physics hypothesis ${\rm NP}_1$ (narrow peak in the tail) for $x\geq0, x>0.3$ and $x>0.5$. No substantial difference is observed in the distributions of the test statistic. As a consequence the expected reach is independent of the cut.}
\label{fig:cuts}
\end{figure}

\subsection{(In-)Sensitivity to Cuts}\label{sec:cuts}

The point is conveniently illustrated in the NP$_1$ example. Since the signal is sharply localized at $x=0.8$, one might expect that restricting the analysis to events in the tail of the distribution, for instance to those with $x>0.3$ or $x>0.5$ will give us a better reach. This would have indeed been the case for the goodness-of-fit test. Our method is instead insensitive to the cut, as figure~\ref{fig:cuts} shows. 

The median significance is $3.1\,\sigma$ for both $x>0.3$ and $x>0.5$. Also the $Z_{\rm{id}}$-$Z$ correlation plot that we do not show here is essentially identical to the one without cut displayed in figure~\ref{fig3}. These results have been obtained using the same procedure outlined in the previous section for the case without cut on $x$. We employed the same learning rate, training algorithm, number of training rounds and network architecture (a single hidden layer with four neurons). The only change is in the number of expected events. However notice that we were not conceptually obliged to choose the same hyperparameters as in the no-cut case. In particular the smaller number of events might have suggested using a smaller network.  It is encouraging that a selection cut does not improve the significance. If our method had been sensitive only in signal-enriched regions ($x>0.5$ for example, where $S/B\approx 0.3$) we would have not solved the problems that plague the binned histogram test, discussed in section~\ref{sec:bi}. Suppose, for concreteness, that we had analyzed data in the $x>0.5$ search region, finding a considerable tension with the reference model. The immediate question, related to the look-elsewhere effect~\cite{Patrignani:2016xqp}, would be whether adding data in the $x\in[0,0.5]$ region would wash out the tension or not. We verified that in our examples this would not be the case, on average, even if new physics is only present at $x>0.5$. Enlarging the search region to the full $x\in[0,1]$ range would at most increase the tension, giving us sensitivity to the possible presence of new physics (such as for instance NP$_3$) that does not show up in the restricted data set.

\subsection{Two Dimensions}\label{2d}

We now consider a $2$-dimensional random variable $x=(M,c)$, with $M\in [0,1]$ and $c\in [-1,1]$. The variable $M$ is interpreted as the invariant mass, while $c$ is the cosine of the scattering angle in the center of mass frame. These two variables conveniently characterize $2$-body final states in LHC events. The distributions of $M$ are chosen among the ones that we previously introduced in the univariate examples. Namely, in the reference model $M$ is exponentially distributed as in eq.~(\ref{SMref1}), while the putative new physics signal is the resonant peak in eq.~(\ref{BSM1}), duly combined with the background as in eq.~(\ref{BSM1prime}). The variable $c$ is uniformly distributed both in the reference and in the new physics model, hence it possesses no discriminating power. This setup makes the comparison between $1$D and $2$D performances particularly meaningful and straightforward. The results obtained in the previous section can indeed be regarded as those that we have if the $2$-dimensional data set is analyzed with the prior bias that $M$ is the only relevant variable. The present section instead discusses what we can get without this prior.

\begin{figure}[!t]
\centering
\includegraphics[width=0.45\textwidth]{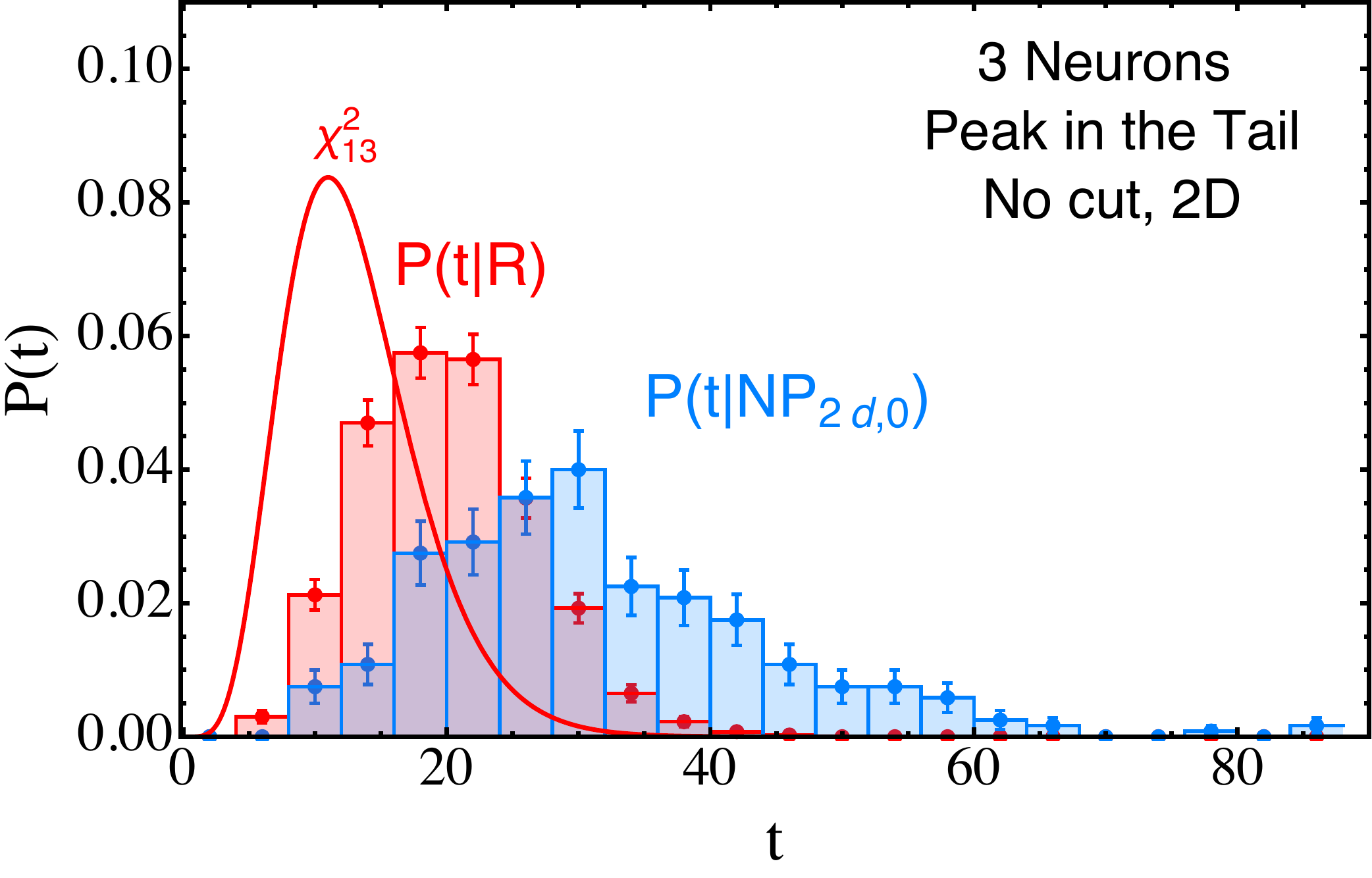}
\caption{Test statistic distribution in the NP$_{2d,0}$ new physics model, compared with the reference one. We expect $2010$ events in the new physics model as in the one-dimensional case.}
\label{fig:2DNP1}
\end{figure}

The test statistic distributions are reported in figure~\ref{fig:2DNP1}. The results are obtained with a $(2,3,1)$ network, trained with the same initial learning rate, training algorithm and training rounds as before. A considerable loss in sensitivity is observed in comparison with the $1$D case in figure~\ref{fig2}. The significance rarely reaches $3\,\sigma$, and the median is $1.4\,\sigma$. The correlation between $Z$ and $Z_{\rm{id}}$ is less sharp, and large-$Z_{\rm{id}}$ samples often end up having low significance. This results from the combination of two distinct effects. The first one is that the values of $t$ resulting from the neural network training on new physics samples are significantly smaller, the second is that $t$ is larger on the reference samples. Let us discuss the two effects separately. 

The new physics median $t$ is now $29$, while it was $36$ in 1D. This result might seem inconsistent, in light of the fact that the 2D network for $M$ and $c$ contains configurations, obtained by setting to zero all the weights for $c$, that are fully equivalent to a 1D network for $M$. However the 1D network obtained in this way has a $(1,3,1)$ architecture, while a $(1,4,1)$ network is employed in figure~\ref{fig2}. A $(1,3,1)$ network in 1D, discussed in the next section, indeed produces a median new physics $t$ of $31$, very close to the 2D one.\footnote{In one dimension the smaller new physics $t$ for the $(1,3,1)$ network does not result in a degradation of the sensitivity because the reference model $t$ distribution is also shifted to lower values, as discussed in the next subsection.} Therefore the new physics median $t$ we find in 2D is not in sharp contradiction with 1D results. Still it is somewhat surprising that it is not larger than the 1D one because the weights associated to $c$ should in principle allow to find a deeper minimum for the loss function.
 This is what happens on reference model samples, whose 2D distribution is shifted to much higher value than those in 1D for the $(1,3,1)$ network (see figures~\ref{fig:2DNP1} and~\ref{fig:architecture}). 

The reference model $t$ distribution is not only shifted with respect to the $(1,3,1)$ network, which follows a $\chi^2$ with 9 degrees of freedom, but also with respect to the $\chi^2_{13}$, in spite of the fact that the $(2,3,1)$ network that we employed has 13 free parameters. We further elaborate on this point in section~\ref{hyp} and in the conclusions. 

The result indicates that improvements in the implementation of our method can be made before considering applications to multivariate data sets. There are many possible directions of investigation in terms of training algorithm and network architecture that we believe would improve the sensitivity in higher dimensions. We discuss them in the Conclusions. However even with this loss in sensitivity, our method should still be explored as a viable alternative to binned histogram model-independent searches which are dramatically affected by the curse of dimensionality.

Furthermore the concrete impact of the  loss in significance that we observe should not be overemphasized. Even if no significant tension is typically found in the 2D data sets under consideration, the signal could still be discovered by running the experiment longer and collecting more events. With twice more luminosity, i.e. $B=4000$ and $S=20$, we obtain a median significance of $2.3\,\sigma$. 

\begin{figure}[!t]
\centering
\includegraphics[width=0.95\textwidth]{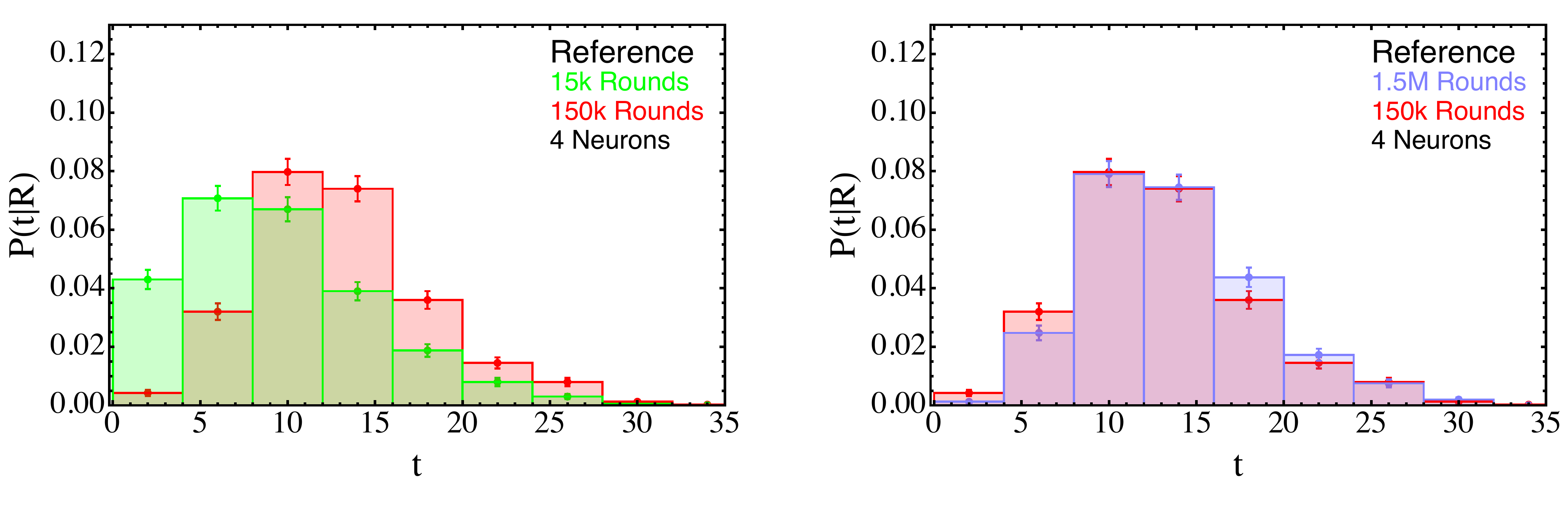}\hspace{15pt}
\caption{Test statistic distribution in the reference hypothesis, $P(t|R)$, for networks with one hidden layer and 4 neurons. {\it Left panel}: 15$\times 10^3$ training rounds compared to 150$\times 10^3$. {\it Right panel}: 1.5$\times 10^6$ training rounds compared to 150$\times 10^3$.}
\label{fig:training}
\end{figure}

\subsection{Dependence on Hyperparameters}\label{hyp}

The aim of this section is to illustrate how the performances depend on the algorithm hyperparameters such as the initial learning rate, the number of training rounds and the architecture of the neural network. 

Our method is founded on maximizing a likelihood function proportional to minus our loss function. Therefore the parameters of the training algorithm should be selected as those that produce the smallest loss, and in turn the largest $t$~in eq.~(\ref{tML}). We verified  that lowering the learning rate below our benchmark value of $10^{-3}$ does not increase $t$. For higher values the loss oscillates as training proceeds and it does not converge. Similarly we verified that ten times more training rounds than the $150\,000$ benchmark do not change the performances. Less training instead would be insufficient. This is shown in figure~\ref{fig:training} for reference-distributed data. The same is found with new physics samples. 

\begin{figure}[!t]
\centering
\includegraphics[width=0.95\textwidth]{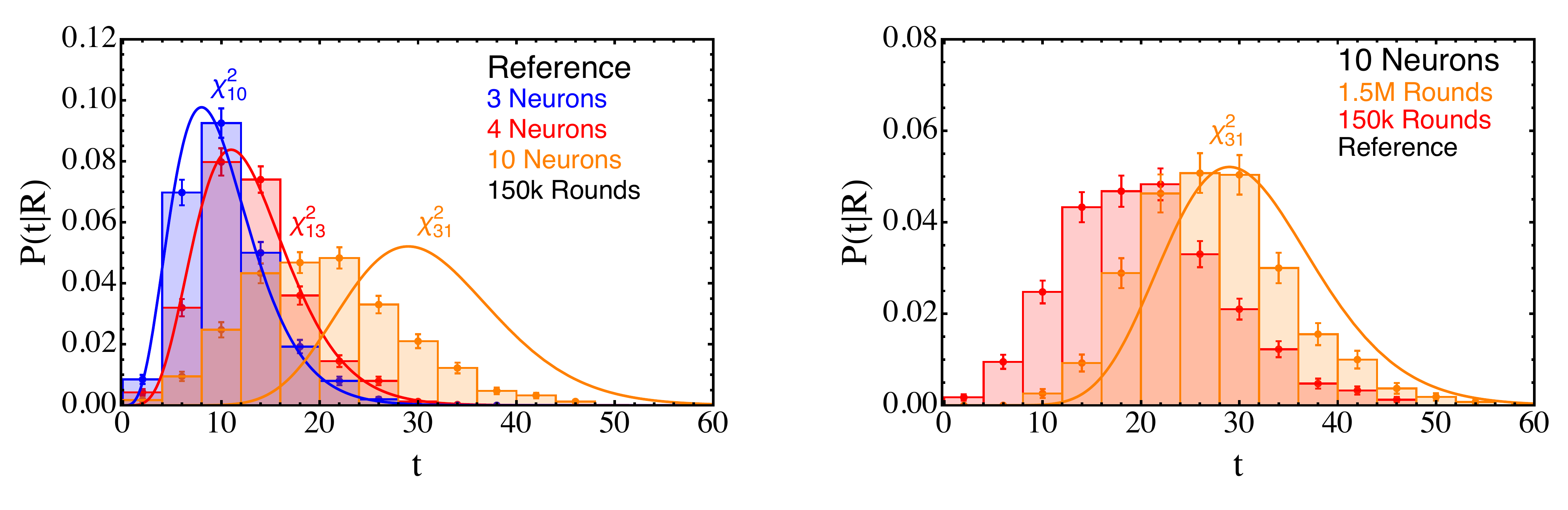}\hspace{15pt}
\caption{{\it Left panel}: Test statistic distribution in the reference hypothesis, $P(t|R)$, for networks with one hidden layer and $3, 4$ or $10$ neurons, compared to the $\chi^2$ with the same number of d.o.f. as the network. The training parameters are the same for all architectures (15000 training rounds, $0.001$ initial learning rate, \textsc{RMSprop} algorithm). {\it Right panel}: Test statistic distribution in the reference hypothesis for the network architecture with 10 neurons. We compare the result with 150 thousands and 1.5 million training rounds. The figure shows how our networks reproduce the asymptotic formulas for the test statistic expected from the theorems in~\cite{Wilks:1938dza, Wald1943}. However larger networks require more training rounds.}
\label{fig:architecture}
\end{figure}

The situation is more interesting if we vary the network architecture. In the left panel of figure~\ref{fig:architecture} we show how the test statistic distribution in the reference hypothesis changes with the number of neurons, while keeping the number of training rounds fixed at $150\,000$. As we increase the free parameters in the network, $t$ increases. 
This has to be expected in light of the well-known result by Wald and Wilks~\cite{Wilks:1938dza, Wald1943} (see also \cite{Cowan:2010js} for a more modern discussion), according to which the maximum log-likelihood ratio test statistics is distributed in the asymptotic limit as a  $\chi^2$ with a number of degrees of freedom which is equal to the number of free parameters in the maximum likelihood fit.\footnote{We are of course referring to the case in which the data are distributed according to the hypothesis that is being tested, i.e. the reference hypothesis in the present case.} In our case the free parameters (i.e., ${\mathbf{w}}$ in eq.~(\ref{tML})) are $10$ for the $(1,3,1)$ network, $13$ for the $(1,4,1)$ network and $31$ in the $(1,10,1)$ case. The $(1,3,1)$ and $(1,4,1)$ distributions follow the asymptotic formula with the corresponding number of parameters, while the $(1,10,1)$ distribution is slightly below the expectation. However this is most likely due to insufficient training. With $1.5$ million training rounds the $(1,10,1)$ distribution tends to align with the $\chi_{31}^2$, as shown in the right panel of figure~\ref{fig:architecture}. \footnote{More training rounds do not change the distribution for smaller networks, as previously mentioned.} The limited computing power at our disposal and the need to perform the training thousands of times on toy data sets did not allow us to check if an even longer training would take the $(1,10,1)$ distribution even closer to $\chi_{31}^2$.

It should be noticed that the asymptotic formulas only hold in the formal limit of infinite statistics, and there are no sharp criteria to establish how many events are concretely needed for them to apply. Therefore the agreement we observe is not a consistency check. It simply means that the statistics in our 1D example is sufficient, at least for networks with up to $10$ neurons, to reproduce the asymptotic distribution. It is legitimate to expect departures from the $\chi^2$ for much larger networks. However we could not verify this fact because the required training time increases with the network capacity, as we have seen. Departures from the $\chi^2$ formula were instead found in the 2D example, see for instance figure~\ref{fig:2DNP1} and section~\ref{2d}. We discuss in the conclusions why it would be important to develop an understanding of this difference between the 1D and the 2D examples.

More concretely, we are interested to know how the sensitivity of the test depends on the neural network architecture. We find that $t$ increases with the network capacity also for new physics generated samples. The median $t$ in the data samples is 31 for the $(1,3,1)$ network, 36 for $(1,4,1)$ and 56 for $(1,10,1)$. This compensates for the growth of $t$ in the reference model, making the significance roughly invariant. We find a median significance of $3.2\,\sigma$, $3.1\,\sigma$ and $3\,\sigma$ for the 3,4 and 10-neurons networks, respectively. Notice however that $1.5$ million training rounds have to be employed in the $10$ neurons case, making the algorithm $10$ times slower. With $150\,000$ rounds we would have obtained a slightly lower significance of $2.7\,\sigma$.

\section{Alternative Loss Functions}\label{sec:altloss}

In sections~\ref{sec:bi} and \ref{sec:alg} we constructed our algorithm as a straightforward application of the maximum likelihood method. Here we describe an alternative derivation, slightly less direct and conceptually rewarding, which however offers more freedom in the implementation. In particular, it allows us to employ different loss functions than the one in eq.~(\ref{eq:loss}). The starting point is the definition of $t$ in eq.~(\ref{tNP}), which we rewrite below for convenience
\beq\label{tNP1}
\displaystyle
t({\mathcal{D}})=2\,\log\left[\frac{e^{-N({{\mathbf{\widehat{w}}}})}}{e^{-N({\rm{R}})}}\prod\limits_{x\in {\mathcal{D}}}\frac{n(x|{{\mathbf{\widehat{w}}}})}{n(x|{\rm{R}})}\right]
\,.
\eeq
This equation instructs us to construct the test statistic as the log ratio between the reference distribution and the ``best fit'' distribution $n(x|{{\mathbf{\widehat{w}}}})$, obtained from the data set under consideration. In eq.~(\ref{tNP}) we are using as best fit distribution the one that maximizes the likelihood (this is why we could add the second equality and express $t$ as the minimum of the likelihood ratio). However eq.~(\ref{tNP1}) still defines a viable test statistic even if we employ a different method to estimate $n(x|{{\mathbf{\widehat{w}}}})$.

Neural network estimators of $n(x|{{\mathbf{\widehat{w}}}})$, or equivalently of $f(x;{{{\mathbf{\widehat{w}}}}})$, can be obtained using different loss functions, the one in eq.~(\ref{eq:loss}) being only one of many possibilities. The loss function that is most widely employed in classification problems is the so-called ``cross-entropy''
\bea\label{eq:lossother}
\displaystyle
&&L[f]=
\sum\limits_{(x,y)}\left[y\,\log[1+e^{-f(x)}]+(1-y)\frac{N({\textrm{R}})}{{\mathcal{N}}_{\mathcal{R}}}\log[1+e^{f(x)}]
\right]\nonumber\\
&&\hspace{23pt}=\sum\limits_{x\in{\mathcal{D}}}\log\left[1+e^{-f(x)}\right]+\frac{N({\rm{R}})}{{\mathcal{N}}_{\mathcal{R}}}\sum\limits_{x\in{\mathcal{R}}}\log\left[1+e^{f(x)}\right]\,.
\eea
The reason why this is a viable choice can be easily understood as follows. In the asymptotic limit, i.e. when the data and the reference sets are large, the sums in eq.~(\ref{eq:lossother}) approach expectation values over the variable $x$. The distribution of the reference sample ${\mathcal{R}}$ is $n(x|{\rm{R}})$ by construction. The data sample ${\mathcal{D}}$ is instead distributed according to the ``true'' data distribution $n(x|{\rm{T}})$, which is precisely the one we would like to estimate. Eq.~(\ref{eq:lossother}) thus approaches the functional
\beq
L[f]\simeq\int dx\,n(x|{\textrm{T}})\log\left[1+e^{-f(x)}\right]+\int dx\,n(x|{\textrm{R}})\log\left[1+e^{f(x)}\right]\,,
\eeq
Let us now take the limit in which the neural network is very large, such that $f(x,{\mathbf{w}})$ effectively spans the whole set of infinitely differentiable functions of $x$. In this limit the minimum of $L[f]$ is where the functional derivative $\delta L[f]/\delta{f}$ vanishes. Therefore  the neural network trained with the loss function in eq.~(\ref{eq:lossother}) is approximately
\beq\label{eq:appr1}
f(x,{{{\mathbf{\widehat{w}}}}})\simeq \log\left[\frac{{{{n}}(x|{\rm{T}})}}{{{{n}}(x|{\rm{R}})}}\right] \,.
\eeq
Since $f(x,{{{\mathbf{\widehat{w}}}}})$ provides an approximation of the true data distribution, it can be meaningfully used to construct the test statistic. Notice that now $t$, unlike in the maximum likelihood approach (\ref{tML}), cannot be directly obtained from the value of the loss function at the end of training. On the contrary it must be evaluated from the definition in eq.~(\ref{tNP1}), using the trained neural network $f(x,{\widehat{{\mathbf{w}}}})$ and evaluating separately the integral of eq.~(\ref{totexp}). This is done with the Monte Carlo method
\beq
N({{\mathbf{{w}}}})=\frac{N({\rm{R}})}{{\mathcal{N}}_{\mathcal{R}}}\sum\limits_{x\in{\mathcal{R}}}e^{f(x,{{{\mathbf{\widehat{w}}}}})}\,,
\eeq
using the same reference sample that is employed for training.

Similar considerations hold for other loss functions such as the square loss or, of course, the maximum likelihood loss in eq.~(\ref{eq:loss}). All of them approach, in the asymptotic limit, integral functionals whose minima give eq.~(\ref{eq:appr1}). Choosing one or the other is from this viewpoint merely a matter of technical convenience. We explored quite extensively the possibility of using the cross-entropy loss. This was actually our first attempt, which we eventually abandoned in favor of maximum likelihood, that was found to have better performances in all the examples we studied. At the technical level the advantage of maximum likelihood is that the test statistic is directly related with the minimum of the loss function. We have seen that this is not the case for other choices of the loss function, hence there is a much less direct connection between $t$ and the quantity that is minimized by the training algorithm. 

Maximum likelihood is normally considered to be the optimal hypothesis test, in accordance with our findings. However it should be kept in mind that for composite alternative hypotheses there is no rigorous notion of optimal test \cite{10.2307/91247}. 

In spite of the fact that maximum likelihood was eventually found to be more effective, the possibility of employing other loss functions should be kept in mind for further evolutions of our algorithm, or for different applications. For instance, we mentioned that another possible application of our method could be the comparison between two samples obtained with different Monte Carlo generators. Since in this case there is no sharp notion of which one is the ``data'' and which one is the ``reference'' sample, one could argue in favor of a more symmetric loss function such as the cross-entropy or square loss. This is left to future work.

\section{Conclusions and Outlook}\label{conc}

We studied the possibility of using neural networks to identify data departures from the prediction of a given reference model, making effectively no assumption on the alternative model that is responsible for the discrepancy. A concrete implementation of the idea was presented, in the form of an algorithm that straightforwardly follows from the maximum likelihood hypothesis test. The inputs of the algorithm are the data collected by an experiment and a reference sample that follows the reference model distribution. The reference data set can be obtained from a Monte Carlo event generator or from data in a control region. Its double role is to replace the analytical knowledge of the reference model distribution, which is typically not available, and to turn likelihood maximization into a supervised training process. The output of the algorithm is the ratio between the best-fit data distribution and the reference one, and a test statistic variable $t$. The former can be used to select data that display the highest level of discrepancy with the reference model. The latter measures the disagreement between the reference model and the data and it can be used for an hypothesis test. 

We performed simple numerical experiments to assess the virtues of our construction and its limitations. We confirmed the model-independent nature of our method, by showing that it has good sensitivity to different hypothetical new physics signals. We also verified that our method does not suffer from the presence of data that agree well with the reference model prediction, even if those constitute the vast majority of the sample. For the applications that we have in mind, as explained in the Introduction and in section~\ref{sec:NNMI}, this is an essential property. Finally we found that the sensitivity does not depend much on the capacity of the neural network. The results above are obtained in a few simple, one-dimensional, examples. A more extensive investigation would be useful to put them on firmer ground.

We also quantified the sensitivity degradation due to including in the network input an additional variable that does not possess discriminating power between the reference and the new physics models. Some amount of degradation is unavoidable, however the one we observed does not reflect the full potential of our approach. On the other hand the sensitivity scales well with the statistics, by doubling the number of events we recover a sensitivity that is comparable to the one dimensional case. Even at fixed number of events we are confident that the situation can be improved by refining our approach. This belief is motivated by the fact that the sensitivity loss in two dimensions comes from a significant departure, towards larger values, of the reference model $t$ distribution with respect to the $\chi^2$ prediction. We do not have a complete understanding of this phenomenon, but we conjecture that it is due to overfitting and to a non-optimal choice of the neural network architecture. Overfitting could be the explanation because it produces bumps and other sharp features that contribute significantly to $t$, which are due to few events that happen to be concentrated in some region of the phase space. Since they result from few events, these contributions to $t$ can violate the asymptotic formula. The behavior is observed in two dimensions and not in one because two dimensional data are much more sparse, hence easier to overfit. If rather than a fully connected $(2,3,1)$ network we had employed an architecture where the variable $c$ has less links than the variable $M$, the performances on the example discussed in section~\ref{2d} would have clearly been better. One might consider the limiting situation where all weights that connect $c$ to the network are set to zero, effectively going back to the one-dimensional $(1,3,1)$ network for which good performances were observed in section~\ref{hyp}. At present it is unclear that this observation could be turned into a systematic optimization strategy. However we notice possible connections with the problem of identifying and eliminating the redundant parameters of a neural network, which goes under the name of ``compression'' in Machine Learning literature \cite{Han2015}.

Another direction of investigation is related with the alternative viewpoint on our approach that we discussed in section~\ref{sec:altloss}. What we are doing is learning from the data a likelihood ratio. We then use it to construct the test statistic. Whether or not the likelihood ratio is learned using  the maximum likelihood loss function is irrelevant from this viewpoint. This suggests that we should look for synergies with recent works \cite{Cranmer:2015bka,Brehmer:2018kdj,Brehmer:2018eca,Brehmer:2018hga} where the problem of approximating likelihood ratios with neural networks has been studied. These studies could also help to model the systematic uncertainties of the reference Monte Carlo, through the formalism of nuisance parameters. We argued in section~\ref{sec:alg} that the problem of systematics is orthogonal to the one that we are addressing, and that it could be solved with standard tools. However studying its interplay with what we are doing would clearly be an important step.

At the purely computational level, the limiting factor of our algorithm is the training time. This can be considerable because we have employed a large number of reference data for training, typically $100$ times the actual data. However one could try to employ the reference sample more efficiently. When we write $N({\bf w})$ as in eq.~(\ref{totexp}) we are effectively using the most naive Monte Carlo integration strategy, more refined techniques might give the same accuracy with much smaller reference samples. For instance one might employ weighted events, obtained by binning the large original reference sample. If the binning is compatible with the resolution on $x$, and in turn with the weight clipping of the neural network, eq.~(\ref{eq:MCint}) could be evaluated accurately using hundreds of reference events rather than hundreds of thousands. Clearly the loss function in eq.~(\ref{eq:loss}) should be updated accordingly.

In this paper we exclusively discussed our method as a possible approach to model-independent new physics searches. However other applications could be envisaged. The first one is constructing an automated tool that compares the predictions of different Monte Carlo generators, using one of the two generators as ``data'', and the other as ``reference''. This might allow to identify subtle discrepancies that might instead escape ordinary comparisons based on the inspection of selected variables. Monte Carlo generators comparison is much easier to implement than model-independent new physics searches because the data sample size is easier to increase. One might also consider our approach for data validation algorithms. The goal there is to establish if raw data produced during a certain, relatively short, period of time were collected under appropriate conditions, or if instead a contingent problem occurred in the data acquisition system. One should thus compare them with previously collected data, which might be used as the reference sample. This should be relatively easy to achieve because the data are abundant and because the reference sample is perfect by definition. Hence one would not need to worry about systematic uncertainties in the reference. We believe that these directions deserve further study.

\section*{Acknowledgments:}
We would like to thank M. Pierini and M. Zanetti for collaboration during the early stages of this work. We also thank N. Arkani-Hamed, L. Biggio, V. Hirschi, M. Papucci, L. Rosasco and N. Toro for useful discussions. We would also like to thank T. Cohen for very useful comments on the manuscript. RTD is supported by the U.S. Department of Energy under Contract No. DE-AC02-76SF00515. 

\appendix
\section{A Short Introduction to Neural Networks}\label{app:NN}
As mentioned in section~\ref{sec:bi}, a neural network is a set of functions. In our notation each architecture corresponds to a family of real functions ${\mathcal{F}}_{\vec{a}}=\{f_{\vec{a}}(x;{\mathbf{w}}), \forall\, {{{\mathbf{w}}}}\}$ of the $d$-dimensional variable $x$, labeled by a vector $\vec{a}$ of integers that specifies the neural network. The functions depend on $N_{\textrm{par}}$ real parameters ${\mathbf{w}}$, generically called ``weights'' in what follows. 

This family of functions, i.e. the neural network, is constructed as the composition of elementary blocks, called layers. In our notation, which follows the one of {\textsc{Mathematica}}~\cite{Mathematica}, layers can be either of the element-wise or of the linear type. An element-wise layer applies a scalar function to all the elements of the input vector, producing an output with the same dimensionality as the input. In our implementation all element-wise layers (i.e. all our activation functions~\cite{Goodfellow-et-al-2016,bishop:2006:PRML}) are logistic sigmoids
\beq
\displaystyle
\sigma(z)=\frac1{1+e^{-z}}\,.
\eeq
As the name suggests, a linear layer performs a linear transformation and the dimensionality of its output ($d_{\textrm{O}}$) can be different from the one of the input ($d_{\textrm{I}}$). It can be represented as
\beq
\left[\lambda_{d_{\textrm{O}},d_{\textrm{I}}}(\vec{z}\,)\right]_{\alpha_{\textrm{O}}}=\sum_{\alpha_I=1}^{d_I}w_{\alpha_{\textrm{O}}}^{\alpha_{\textrm{I}}}z_{\alpha_{\textrm{I}}}+{\overline{w}}_{\alpha_{\textrm{O}}}\,, \label{eq:linear}
\eeq
where $\alpha_{\textrm{O}}$ runs from $1$ to $d_{\textrm{O}}$. The free parameters of a linear layer are the $d_{\textrm{O}}$ times $d_{\textrm{I}}$ entries of the $w_{\alpha_{\textrm{O}}}^{\alpha_{\textrm{I}}}$ matrix, plus the $d_{\textrm{O}}$ shifts ${\overline{w}}_{\alpha_{\textrm{O}}}$, for a total of $d_{\textrm{O}}(d_{\textrm{I}}+1)$ parameters. We denote all of them as weights in spite of the fact that the $\overline{w}$'s are often called ``biases'' in the Machine Learning literature.

A neural network is the composition of layers, alternating linear and element-wise ones
\beq
f_{\vec{a}}(\,\cdot\, ;{\mathbf{w}})=\lambda_{a_{\rm\sc{L}}=1,a_{{\rm\sc{L}}-1}}\circ\sigma\circ\ldots\circ\sigma\circ\lambda_{a_2,a_1}\circ\sigma\circ\lambda_{a_1,a_0=d}\,,
\eeq
If the network is fully connected, i.e. the dimension of the output of layer $n-1$ equals that of the input of layer $n$, for every layer, then the total number of free parameters that the network depends on is
\beq
N_{\textrm{par}}(\vec{a})=\sum_{n=1}^{{\rm\sc{L}}} a_{n}(a_{n-1}+1)\,.
\eeq
The neural network function is applied to the variable $x$, hence the input of the first linear layer has dimensionality $a_0=d$. The neural network output that we are interested in must be a real number, hence $a_{\rm\sc{L}}=1$. We are instead free to choose the remaining ${\rm\sc{L}}-1$ entries of the $({\rm\sc{L}}+1)$-dimensional vector $\vec{a}$. Notice that ${\rm\sc{L}}$ only counts the number of linear layers in the network. However we often refer to it as the number of layers, matching in this way the more standard terminology in which one ``layer'' is the composition of a linear transformation with $\sigma$. For instance, a two-layers neural network acting on a one-dimensional input variable $x$ is represented by the vector $\vec{a}=(1,N_{\rm{neu}},1)$, where $N_{\rm{neu}}$ is the number of neurons. 

In eq.~(\ref{eq:linear}) each neuron corresponds to a different value of $\alpha_O$. So $\vec{a}=(1,N_{\rm{neu}},1)$ depends on $3N_{\rm{neu}}+1$ free parameters and its explicit functional form is
\beq
f_{(1,N_{\rm{neu}},1)}(x;{\mathbf{w}})=\sum\limits_{\alpha=1}^{N_{\rm{neu}}}(w_{(2)})^\alpha\sigma\left[
(w_{(1)})_\alpha x+({\overline{w}}_{(1)})_\alpha
\right]+{\overline{w}}_{(2)}\,.
\eeq
For the applications considered in this paper we have employed simple networks of this class. However we have tested also deeper networks (${\rm\sc{L}}>2$) for $d>1$ finding comparable performances. 

Once we have built the network, we need to train it. This is not different from fitting free parameters {\bf w} given experimental observations. In analogy with maximum likelihood parameter estimation, we write down a loss function that at the minimum gives estimators of the values of {\bf w} that best describe the data. Then we need to find the minimum.

The choice of loss function is determined by the specific problem at hand. In section~\ref{sec:alg} we have already discussed what we consider the most motivated construction for our model-independent searches and in section~\ref{sec:altloss} we showed a variation based on more standard classification problems. Here we illustrate the point with a simpler example for the readers that are not familiar with the subject. For concreteness we discuss what one would do for supervised learning and refer the reader interested in semi-supervised, unsupervised and reinforcement learning to~\cite{Goodfellow-et-al-2016, Alom2018, bishop:2006:PRML, Hastie2009}. 

Imagine that you have two sets of pictures one of cats and one of dogs. You would like the network to output $1$ if given a cat and $0$ for a dog. In this case the input $x$ can be an array of numbers, each representing a different pixel of the picture. Then an obvious choice for the loss function would be
\bea
L[f] = \sum_{x \in {\rm cats}} \left[1-f_{\vec a}(x|{\bf w}) \right]^2+ \sum_{x \in {\rm dogs}} \left[f_{\vec a}(x|{\bf w)}\right]^2\, . \label{eq:loss_example}
\eea
At the minimum of $L$,  $f_{\vec a}(x_{\rm cat}|{\bf \widehat w})=1$ and $f_{\vec a}(x_{\rm dog}|{\bf \widehat w})=0$. It is very easy to prove it, by taking a functional derivative of $L$ with respect to $f$. What is actually implemented in a computer consists in taking the derivatives of $L$ with respect to the weights going backwards from the last layer. 

Note that the form of the loss function in~(\ref{eq:loss_example}) is just illustrative. As we have also mentioned in the main body of the text, in practical applications the cross-entropy, the Kullback-Leibler divergence and their variations are more widely used. One quality that they have over the $\chi^2$ used in~(\ref{eq:loss}) is that their logarithms cancel the exponential saturation of sigmoids and hyperbolic tangents at least for the last layer, making the derivatives larger and the minimization process faster for certain values of the input.

Since the loss functions obtained by nesting layers are in general non-convex there are no algorithms that are guaranteed to find a global minimum. The prevailing approach consists in finding a ``good enough" local minimum by using Stochastic Gradient Descent. Gradient Descent simply consist in taking a derivative of the loss function and updating the weights by moving them a small amount $\epsilon$ in the direction in which the derivative decreases. This technique was proposed by Cauchy in 1847~\cite{S2010}. The parameter $\epsilon$ is called learning rate. It can be fixed a priori or changed adaptively during training. 
Since computing the derivative over the entire training sample is usually computationally unfeasible, it is typically computed on a subsample chosen at random. This is what goes under the name of Stochastic Gradient Descent~\cite{Goodfellow-et-al-2016,Alom2018}. The \textsc{RMSprop} algorithm~\cite{RMSprop} that we employ is based on Stochastic Gradient Descent.

The process of evaluating $L$ on a subset of the cats and dogs sample, taking its derivatives and updating the values of the weights is known as training and the sample used for the process is known as the training sample. This comes in as many repetitions as it takes to obtain an acceptable degree of accuracy. The accuracy of classifiers, as the one in this simple example, can be tested on a separate sample, (you guessed it) the testing sample. In the applications discussed in the paper, where we are not solving a classification problem, we can perform a different test, by comparing the neural network estimation of the data distribution with its true functional form.

It can be proven that a function built following the procedure outlined at the beginning of this section, can approximate with arbitrary accuracy any continuous function in a compact domain of $\mathbb{R}^N$. For a more precise statement of the relevant theorems we refer to~\cite{Function1989, Hornik1991,Kreinovich1991, Haykin1998}. Here we would like to present a heuristic argument that will also make clear why neural networks provide a good parametrization for the problem described in this work.

\begin{figure}[!t]
\begin{center}
\includegraphics[width=0.6\textwidth]{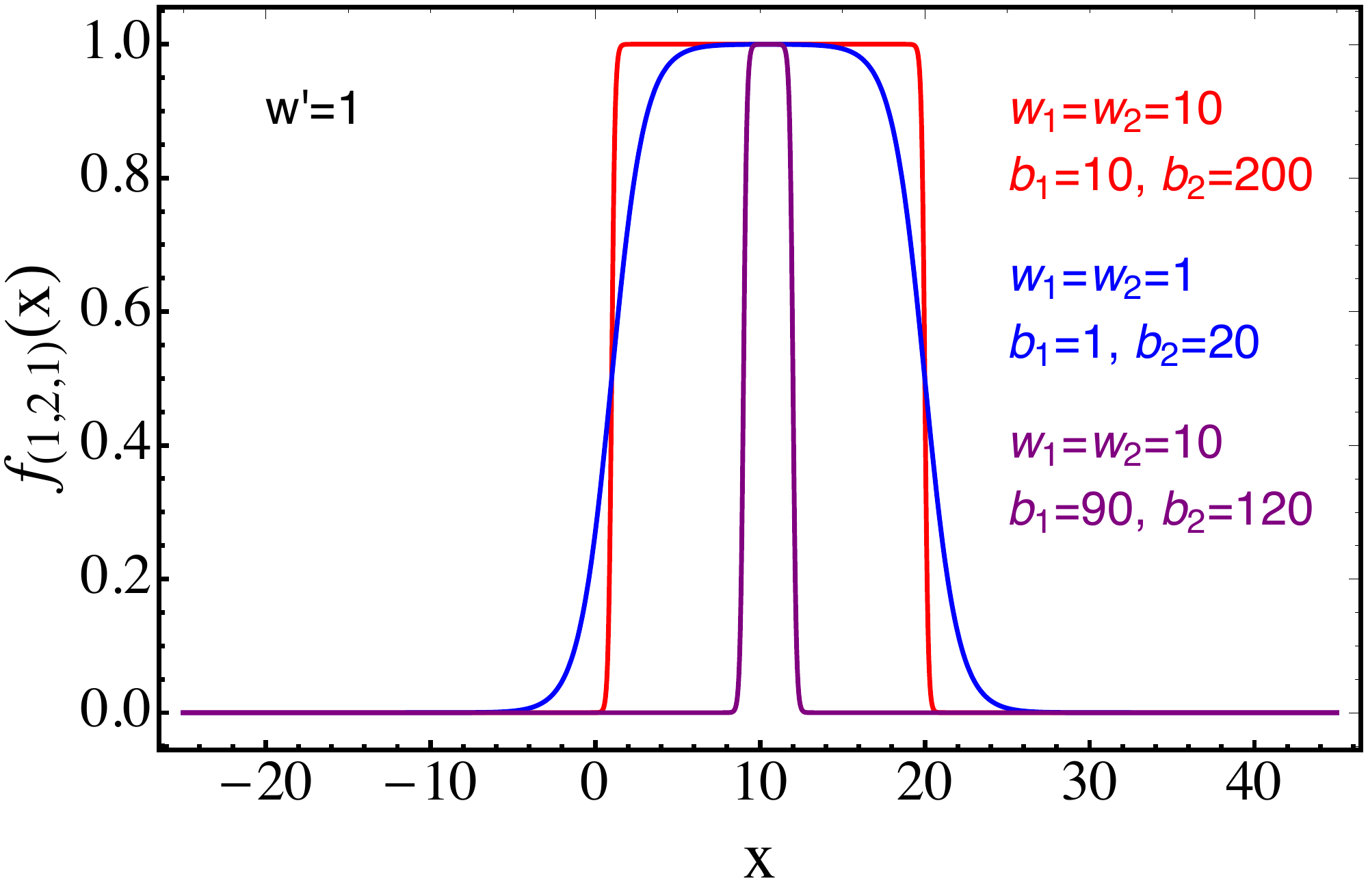}
\caption{Illustration of how three-neurons with logistic sigmoid activation functions can reproduce a rectangular function or a smooth peak. The parameters in the legend of the plot are defined in Eq.s~(\ref{eq:nn1}) and (\ref{eq:nn2}).}
\label{fig:sigmoid}
\end{center}
\end{figure}

Take two neurons with a logistic sigmoid activation function and send their output to a third one. For simplicity consider a one-dimensional input for the first layer. The function that describes this small neural network is
\bea
f_{(1,2,1)}(x) =w_1^\prime \sigma(z_1(x))+w_2^\prime \sigma(z_2(x)) + b^\prime\, , \quad z_i(x) = w_i x + b_i\, \label{eq:nn1}
\eea
where $i=1,2$ labels the two initial neurons. For $w_1^\prime=-w_2^\prime=w^\prime$ and $b^\prime=0$ we have
\bea
f_{(1,2,1)}(x) = w^\prime \left[\sigma(w_1 x + b_1)-\sigma(w_2 x + b_2)\right]\, . \label{eq:nn2}
\eea
This is plotted as a function of $x$  in Figure~\ref{fig:sigmoid}. It is approximately zero for $x\gtrsim -b_2/w_2$ and $x\lesssim -b_1/w_1$ and roughly constant and equal to $w^\prime$ otherwise.

As illustrated in Figure~\ref{fig:sigmoid}, by increasing $w_1$ and $w_2$ we can make the transition between zero and $w^\prime$ arbitrarily sharp. By adjusting $b_1$ and $b_2$ we can make the domain over which $f_{(1,2,1)}(x)$ is non-zero as narrow as we want. So we can make this three-neurons unit generate a smooth peak, a broad plateau or a rectangular function. By combining many of these units we can approximate any continuous function as a juxtaposition of rectangular functions. In higher dimensions we can repeat this argument by adding two more neurons for each new direction. We can send all their outputs into a single final neuron and construct a multidimensional rectangular function in the same way.

As discussed in section~\ref{sec:bi} this also shows why neural networks are promising candidates for new physics searches. Even if we do not know a priori the type of signal that we are looking for, a network with very few parameters can reproduce an arbitrarily sharp feature, remaining smooth in its absence. Fewer free parameters mean a smaller look-elsewhere effect and a larger sensitivity.

\bibliographystyle{utphys}
\bibliography{bibliography}

\end{document}